\documentclass[floatfix,10pt,superscriptaddress,twocolumn,showpacs,showkeys,aps,prb,final]{revtex4-1}

\usepackage{graphicx}
\usepackage{dcolumn}
\usepackage{subfigure}
\usepackage{bm}
\usepackage{ifpdf,braket}
\usepackage{epstopdf}
\usepackage{float}
\usepackage{enumerate}
\ifpdf
\usepackage[pdftex,
        colorlinks=true,
        pdftitle={Excited vortex states in superconductor},
        pdfauthor={Mauro M. Doria},
        baseurl={http://www.if.ufrj.br/~mmd},
        pdftoolbar=true,
        pdfmenubar=false,
        pdfpagemode=UseOutlines,
        pdfstartview=FitH,
        bookmarksopen=false
        ]{hyperref}
\fi

\begin{document}

\title{Multigap superconductivity and interaction driven resonances in superconducting nanofilms with an inner potential barrier}
\author{Mauro M. Doria}
\affiliation{School of Pharmacy, Physics Unit, Universit\`{a} di Camerino, 62032 - Camerino, Italy}
\affiliation{Instituto de F\'{\i}sica, Universidade Federal do Rio de Janeiro, 21941-972 Rio de Janeiro, Brazil}%
\email{mauromdoria@gmail.com}
\author{Marco Cariglia}%
\affiliation{School of Pharmacy, Physics Unit, Universit\`{a} di Camerino, 62032 - Camerino, Italy}
\affiliation{Departamento de F\'{\i}sica, Universidade Federal de
Ouro Preto, 35400-000 Ouro Preto Minas Gerais, Brazil}
\author{Andrea Perali}%
\affiliation{School of Pharmacy, Physics Unit, Universit\`{a} di Camerino, 62032 - Camerino, Italy}
\affiliation{INFN, Sezione di Perugia, 06123 - Perugia, Italy}

\begin{abstract}
We study the crossover in a zero temperature superconducting nanofilm from a single to a double superconducting slab induced by a tunable insulating potential barrier in the middle.
The single phase superconducting ground state of this heterostructure is shown to be intrinsically multigapped and to have a new type of resonance caused by the strength of the barrier, thus distinct from the Thompson-Blatt shape resonance which is caused by tuning the thickness of the film.
Single particle electronic states are strongly or weakly affected according to their parity (even or odd) with respect to the insulating barrier.
The lift of the parity degeneracy at finite barrier strength reconfigures  the pairing interaction and leads to a multigapped superconducting state with interaction driven resonances.
\end{abstract}

\maketitle

\section{introduction}
Since two-gap superconductivity has been discovered in MgB$_2$~\cite{liu01, giubileo01, giubileo02, iavarone02, iavarone05, kortus01, bianconi2007} an intense research activity has been devoted to  multiband and multigap superconductors.
In fact two-gap superconductivity has been theoretically proposed  long ago~\cite{suhl1959,moskalenko1991} and recently found in many compounds~\cite{perucchi12} ranging from composites~\cite{hafiez15} to metallic  Pb~\cite{floris07,ruby15}.
Nano-engineered superconducting films tailored to atomic precision thickness~\cite{guo04,bao05,zhang10} are  pivotal to this research as they display multibands and  multigapped superconductivity.
Superconductivity in ultrathin nanofilms was demonstrated to be very robust and it can survive even in monoatomic layers of In and Pb\cite{zhang10}.

In this paper we show that the superconducting nanofilm with an inner potential barrier has an intrinsic multigap structure.
Interestingly a nanofilm with a single superconducting slab has multibands but only displays multigaps within the narrow window of shape resonances, as shown by Thompson-Blatt~\cite{thompson63}. However this narrow window, defined by the nanofilm width and by the pairing energy scale, renders very difficult its experimental observation.
To overcome this problem we consider here a {\it SIS nanofilm}, a superconductor-insulator-superconductor ``sandwich'' such that the two superconductors are thin slabs in the quantum-size regime separated by a very thin insulating slab.
We show here that the SIS nanofilm is multigapped since it remains so whether or not in a shape resonance.
We also show that the SIS nanofilm possesses a new type of resonance hereafter called {\it interaction driven resonance} to distinguish from the well-known {\it shape driven resonance} occurring in the single slab superconducting nanofilm.
The SIS nanofilm displays very distinct and novel properties with respect to the single slab nanofilm. They stem from the additional parity properties with respect to the insulating barrier since single particle states can be either even or odd. Even states strongly feel the presence of the barrier while odd ones do not and this causes a splitting in the single particle energy levels which is at the heart of the multigapped structure of the superconducting state.

The interaction and the shape driven resonances share a single common origin albeit they have distinct properties.
To understand  their common origin we recall that single particle electronic states in films are a sum over continuum and discrete degrees of freedom, associated to the parallel and  perpendicular to the film surface degrees of freedom, respectively.
The origin of the discreteness is in the quantum size regime that renders the superconducting gap smaller than the splitting between two consecutive energy levels perpendicular to the nanofilm.
The discreteness no longer holds for a thick film since consecutive single particle states become so close in energy that their splitting is smaller than the superconducting gap.
The continuum treatment is always suited for the degrees of freedom parallel to the nanofilm surface, where the separation between consecutive levels is always smaller than the superconducting gap.

Superconducting nanofilms have a far richer structure than  a bulk superconductor, the latter defined here by the simplest possible model, i. e., that of a single spherically symmetric three-dimensional Fermi surface.
Even within this simple description nanofilms are far more complex than the bulk. They display multiple two-dimensional Fermi surfaces, each associated to a distinct discrete state induced by the perpendicular confinement.
Shape and interaction driven resonances are a consequence of these multiple Fermi surfaces in nanofilm that cause a reconfiguration of the pairing interaction.
The intersection of the chemical potential with the parabolic parallel bands defines multiple Fermi surfaces, where an  attractive interaction leads to superconducting gaps around the Debye energy window centered in the chemical potential itself.
A resonance in the superconducting gap is the consequence of the entrance (exit) of a two-dimensional Fermi surface into the Debye energy window.
An intersection with a two-dimensional  Fermi surface can be either added or removed and this affects the superconducting gap of the nanofilm. This phenomenon has no counterpart in the simplest description of the bulk superconductor.
The adjustment of some nanofilm parameters moves the discrete levels and is responsible for resonances.
The Thompson-Blatt superconducting shape resonance~\cite{perali96, shanenko06} is due to a change in the nanofilm thickness that affects the discrete levels.
This is the only way to make a single slab superconducting nanofilm resonate since it lacks any other internal structure.
However the SIS nanofilm has an internal structure given by the insulating barrier and this opens a new venue for resonances through the adjustment of the discrete levels by the barrier strength. As shown here the even discrete levels are very sensitive to the barrier strength and through this mechanism can be moved to enter the Debye energy window and cause a resonance in the superconducting gap.

To unveil the intrinsic multigap structure and the existence of interaction driven resonances in the SIS nanofilm we study its zero temperature properties under a fixed chemical potential.
Indeed the chemical potential can be considered approximately constant in case  the number of atomic monolayers in the superconducting nanofilm surpasses a small number, known to be nearly five according to Fig.~1 of Ref.~\onlinecite{thompson63}.
Such assumptions  add simplicity to our study without compromising the present goals.

In this paper we consider that the SIS nanofilm has only one single superconducting state in thermodynamical  equilibrium.
Cooper pairs tunnel through the insulating barrier and there is no phase difference between the two superconducting slabs and so, there is no Josephson current, which can only exists in case of a phase difference between them~\cite{josephson1962,golubov76}.
Although we compute here excited states (multigaps) of the SIS nanofilm those are assumed to be translational invariant along the film surface and so there is no spontaneous current passing from one side to the other of the SIS nanofilm.
Thus Josephson vortices are excluded as those induce localized circulating currents from one slab to the other.

We use the Bogoliubov de Gennes (BdG) equations in the Anderson approximation to show that superconductivity is  multigapped and also to show the existence of interaction driven resonances in the SIS nanofilm.
We do it in the simplest possible theoretical framework able to capture the key elements of the SIS nanofilm.
The  insulating barrier is treated as a repulsive delta function potential, which is enough to describe the parity breaking in the single particle electronic states that can be either affected (even) or not (odd) by it.
The delta function description has the advantage of keeping the parabolic nature of bands even in presence of the insulating barrier.
We find the present simple approach useful with respect of more elaborate treatments~\cite{kim04,serrier13} that can deal with features such as the proximity effect in the SIS nanofilm.
A recent detailed investigation of the shape resonances at the critical temperature
for a single superconducting nanofilm has been reported in Ref. \onlinecite{valentinis16a,valentinis16b}, by an exact numerical solution
of the BCS mean-field equations at fixed density. The main outcome is that
the precise form of the shape resonance and the enhancement (or suppression)
of the critical temperature depend on the strength of the confining potential at the nanofilm surface and on the value of the pairing interaction.
The effects of interaction driven resonances on the critical temperature of a SIS nanofilm will be considered elsewhere.

From the experimental point of view there are several ways to realize the insulating barrier of the SIS nanofilms. One possibility is to deposit on the Nb nanofilm a thin film of Al and then
proceed with the insitu oxidation of Al at room temperature. The thickness of the AlO$_x$ insulating barrier will change depending on the oxygen exposure time, allowing for a control of the barrier potential strength \cite{russo14}.
AlO$_x$ barriers with thickness of the order of 1nm should be realizable. Another possibility could be to use silicon as an insulating barrier in the SIS nanofilms.
\begin{figure}[hb]
\begin{center}
\includegraphics[width=0.7\linewidth]{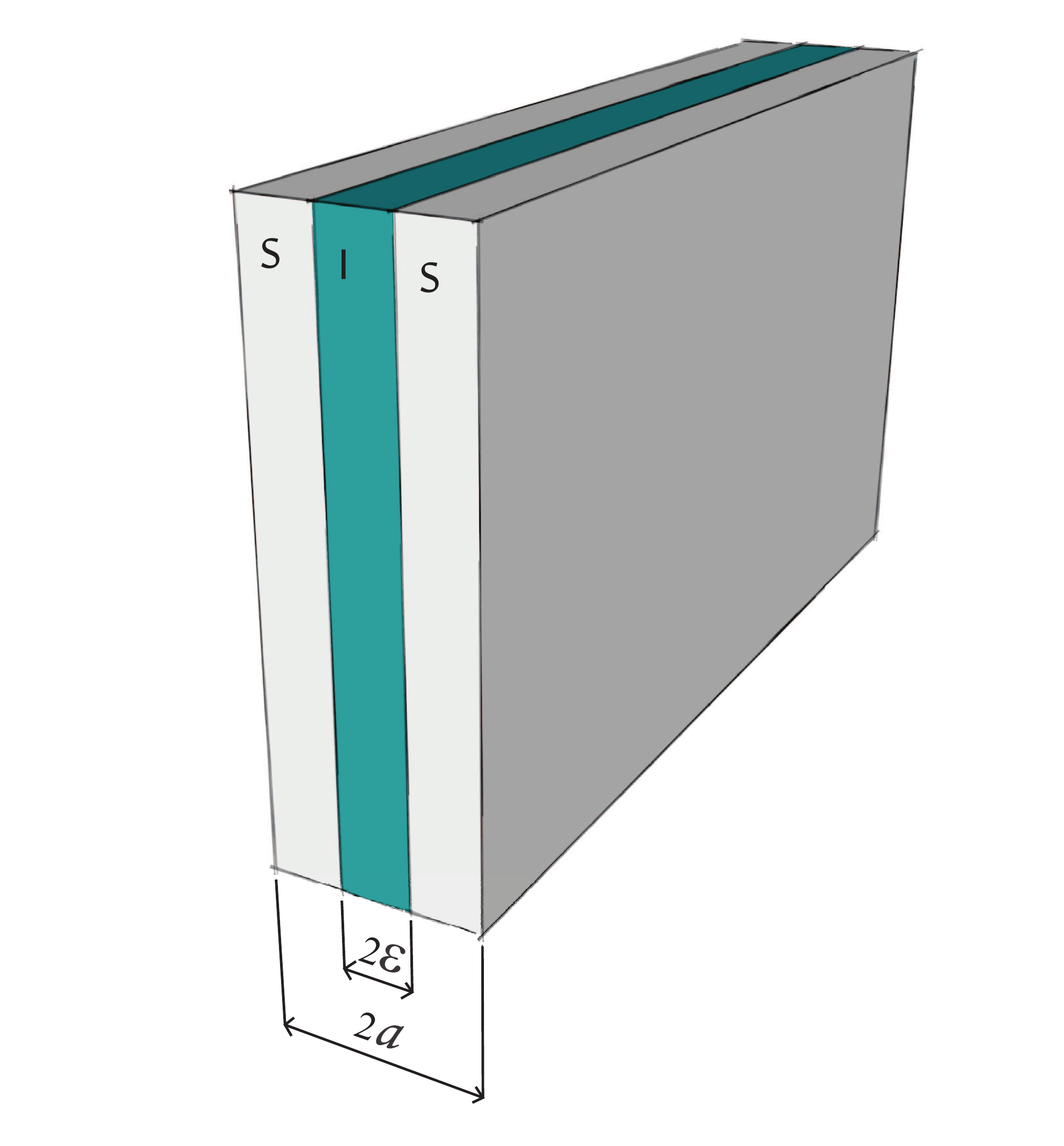}
\caption{(Color on line) A pictorial view of the SIS nanofilm composed of two superconducting slabs with thickness $a-\varepsilon$ and an insulating barrier with thickness $2\varepsilon$ such that the total thickness is $2a$.}\label{sandwich}
\end{center}
\end{figure}
\begin{figure}[hb]
\begin{center}
\includegraphics[width=0.6\linewidth]{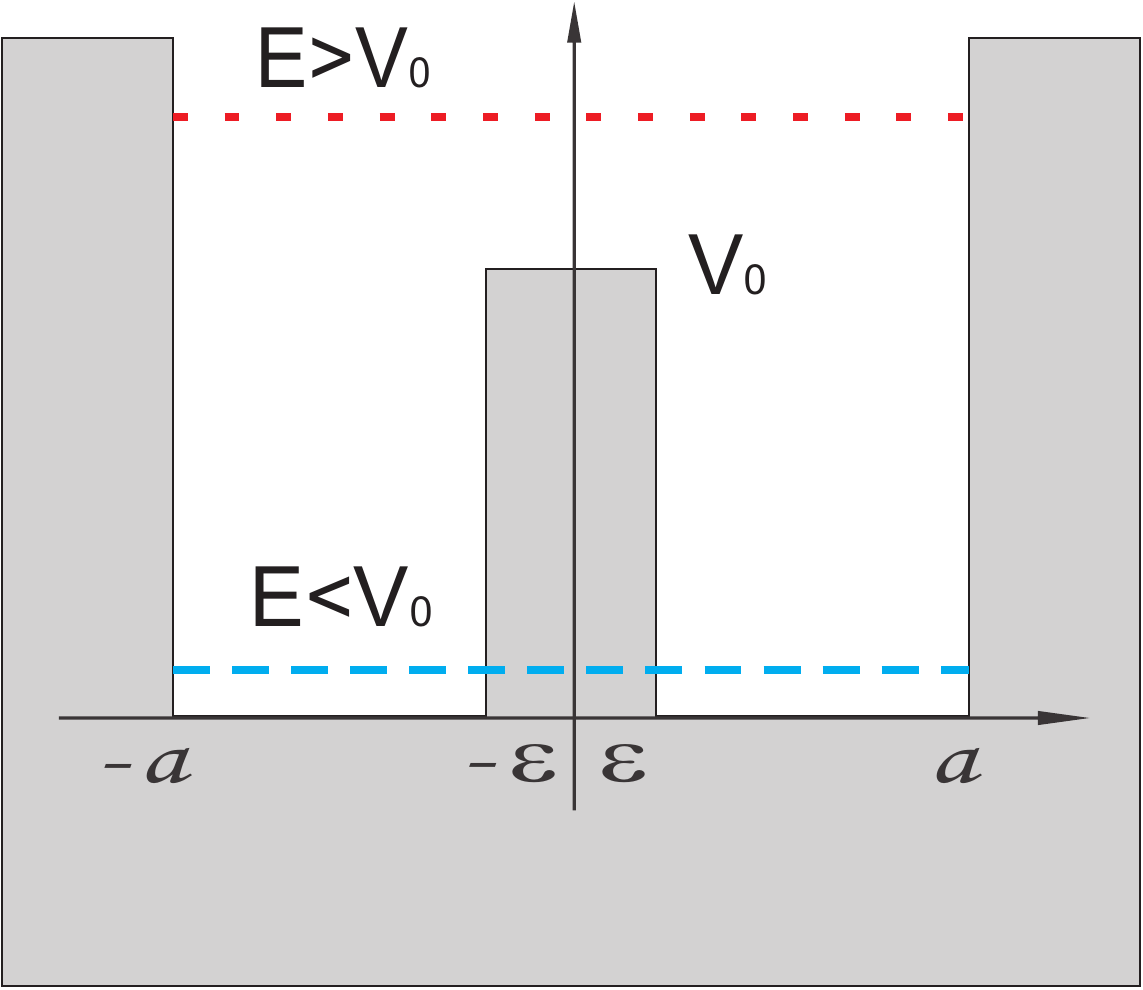}
\caption{(Color on line) Profile of the insulating barrier in the middle of the superconducting nanofilm described by Eq.(\ref{finitebarrier}).
Single particle energy levels above or below the barrier (height $V_0$) are shown here.}\label{pot-well}
\end{center}
\end{figure}
\begin{figure}[hb]
\begin{center}
\includegraphics[width=0.7\linewidth]{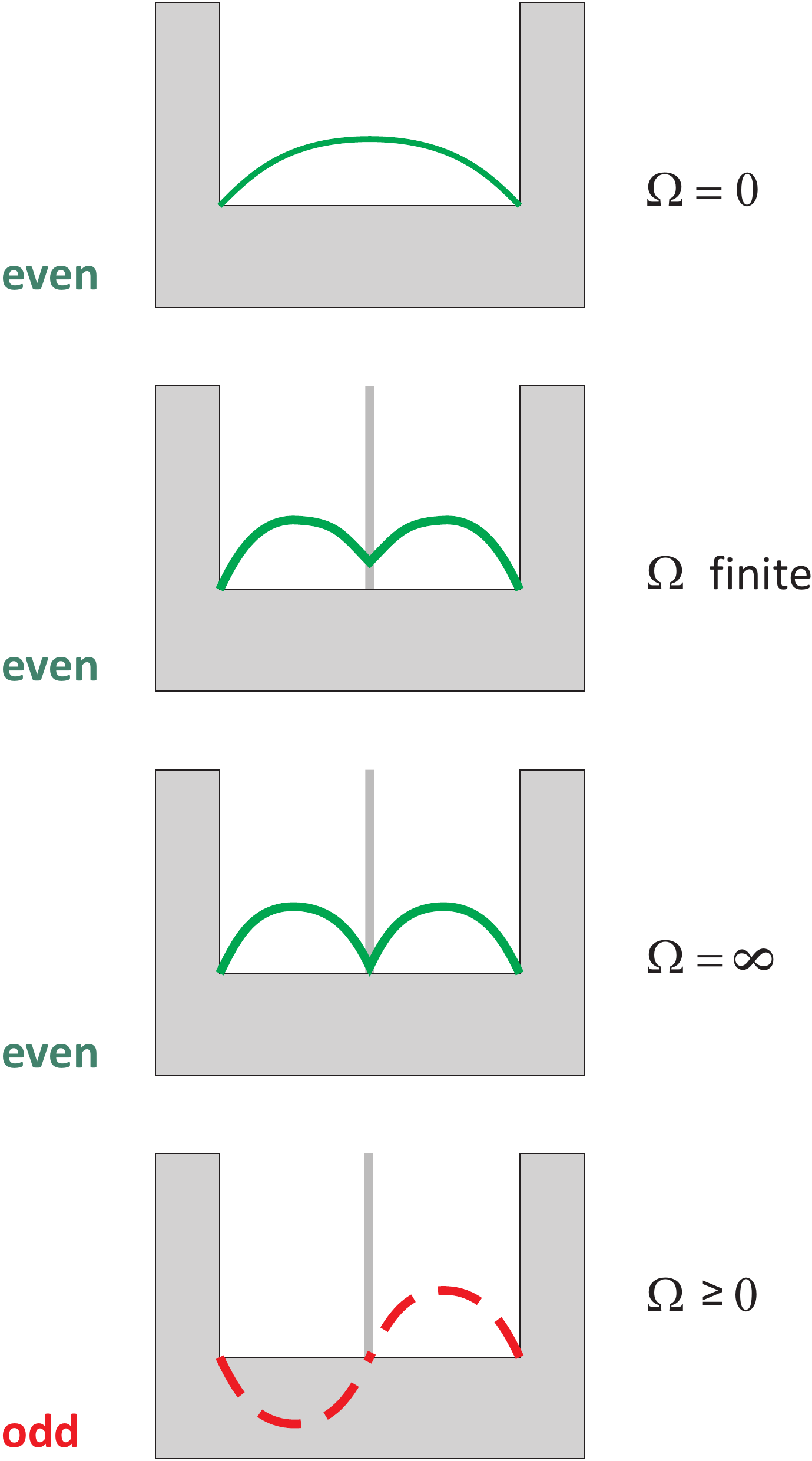}
\caption{(Color on line) The wave function of the even single particle ground state is represented as a function of the barrier strength, $\Omega$, defined in Eq.(\ref{omegadef}).
The first excited state is odd and its wave function remains unchanged for any value of $\Omega$.
The even and odd wave functions  are shown in full (green) and dashed (red) lines, respectively.
}\label{even-odd-vs-omega}
\end{center}
\end{figure}
\begin{figure}[hb]
\begin{center}
\includegraphics[width=0.7\linewidth]{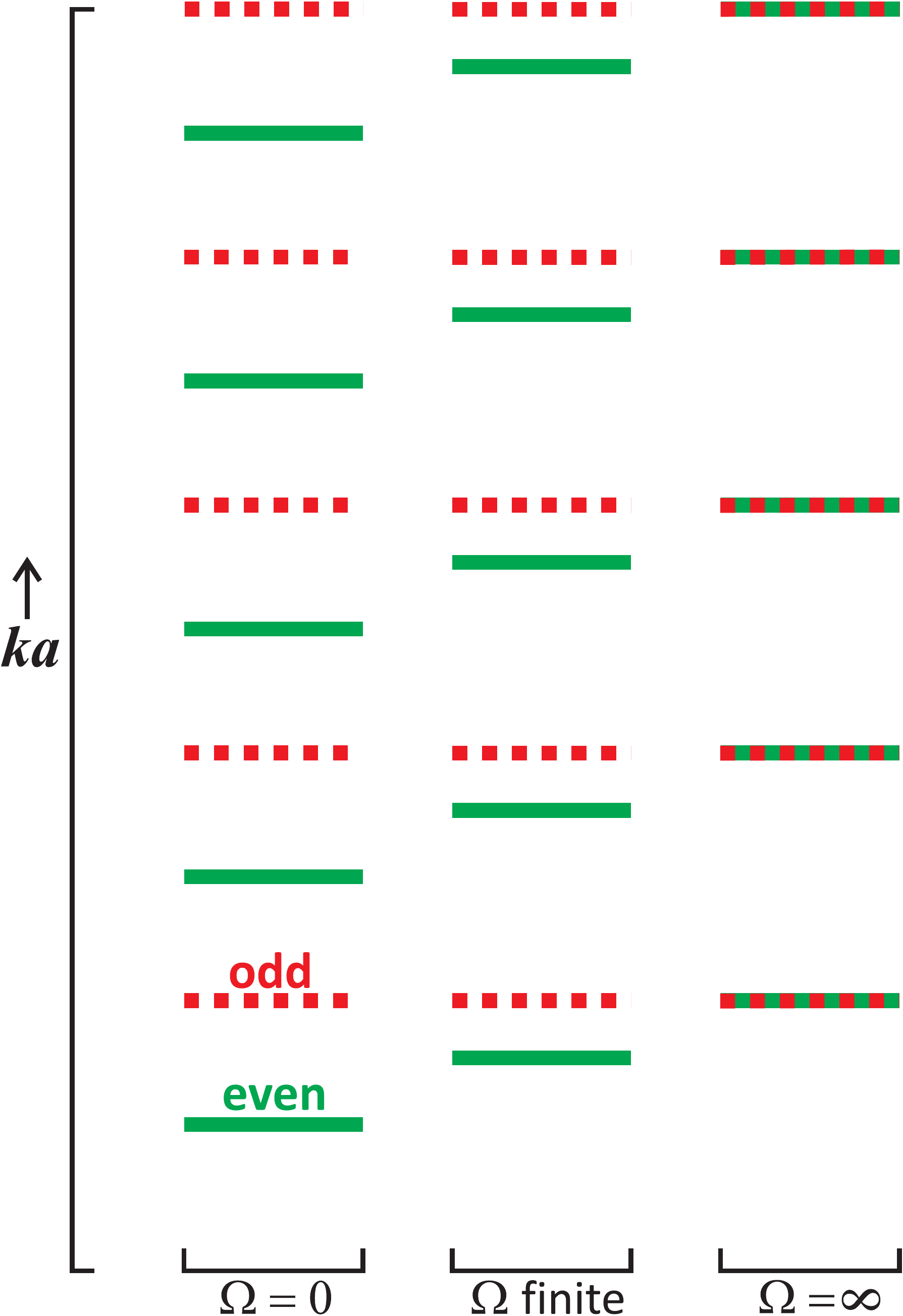}
\caption{(Color on line) Scheme of the single particle energy levels according to their wave number represented by $k a$.
Single particle even and odd states are shown in filled (green) and dashed (red) lines respectively.
Three situations are displayed according to the potential strength defined in Eq.(\ref{omegadef}).
For $\Omega=\infty$ even and odd levels are degenerate.}
\label{levels-vs-omega}
\end{center}
\end{figure}
\begin{figure}[hb]
\begin{center}
\includegraphics[width=1.0\linewidth]{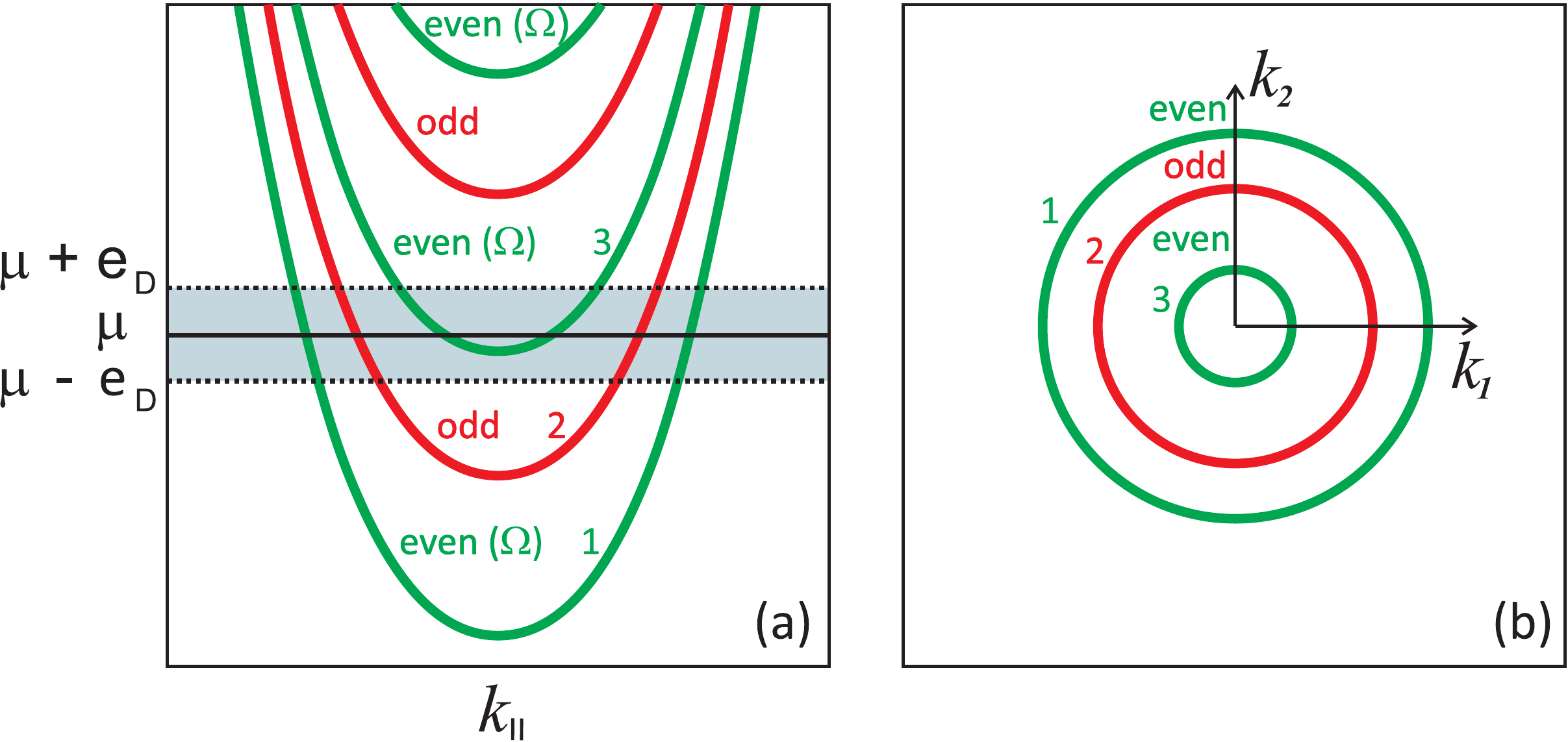}
\caption{(Color on line)
Panel (a) shows the intersection between the parabolic bands, defined by Eq.(\ref{sumener}) and the Debye energy window, defined by Eq.(\ref{debye}).
Panel (b) shows the multiple two-dimensional Fermi levels defined by Eq.(\ref{fermi2d}).
Even and odd states alternate in growing energy and are shown in distinct colors (green and red, respectively).
 }\label{chemical-potential-bands}
\end{center}
\end{figure}
\begin{figure}[hb]
\begin{center}
\includegraphics[width=0.7\linewidth]{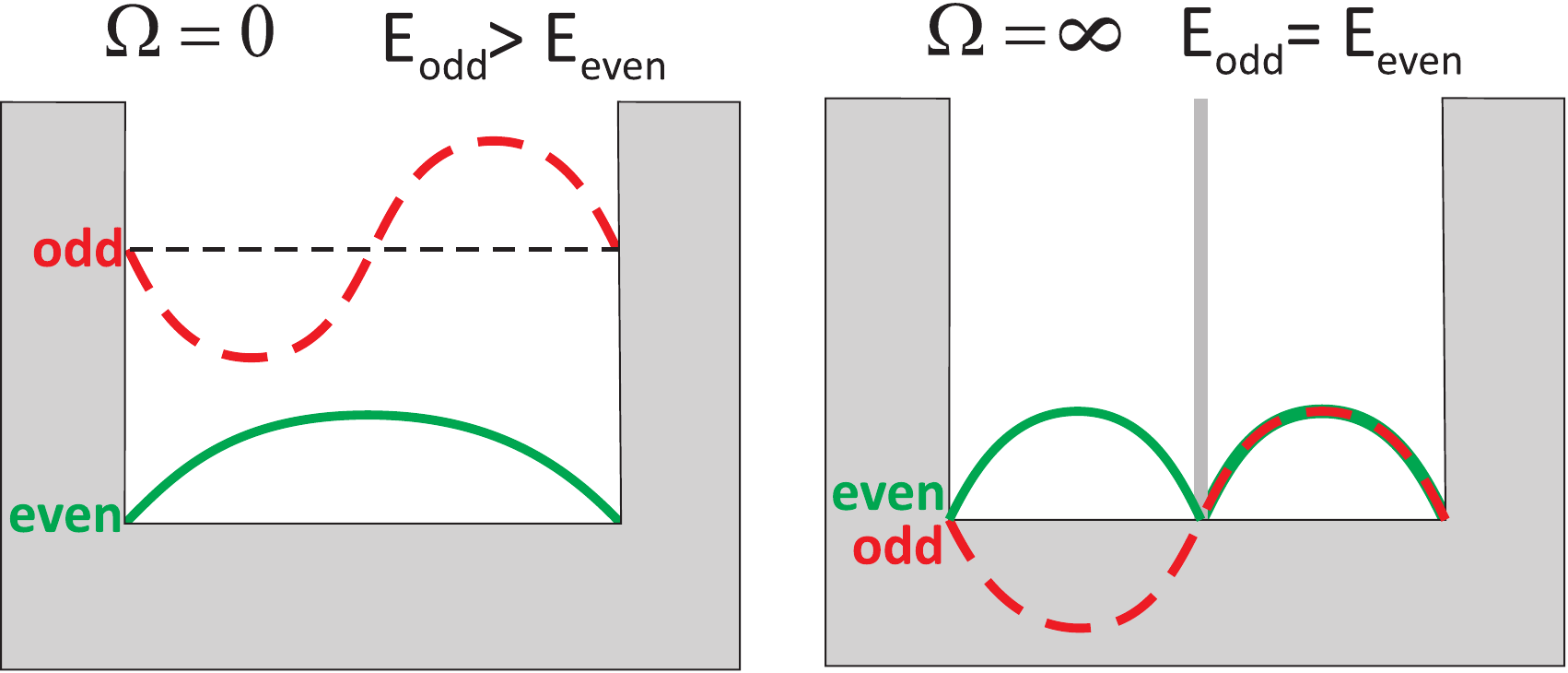}
\caption{(Color on line)
The SIS nanofilm is represented here for the cases of no barrier ($\Omega=0$) and of an infinite barrier ($\Omega=\infty$) which are equivalent to a single and to two decoupled slabs, respectively.
For $\Omega=0$ the single particle ground state is even and the first excited  odd whereas for $\Omega=\infty$ the single particle even and odd states are degenerate in energy.
The even and odd wave functions  are shown in full (green) and dashed (red) lines, respectively.
}\label{pot-zero-infty}
\end{center}
\end{figure}
\begin{figure}[hb]
\begin{center}
\includegraphics[width=1.0\linewidth]{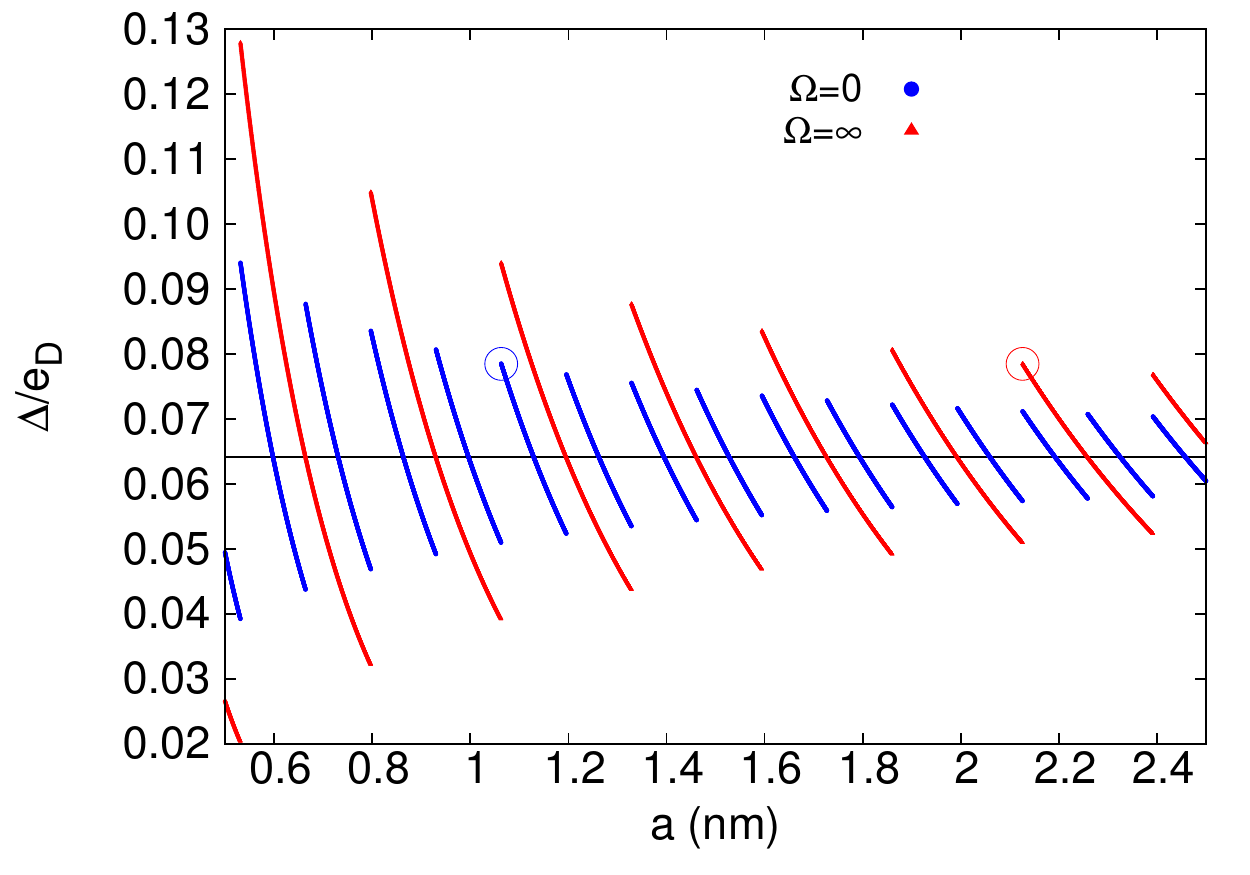}
\caption{(Color on line) The superconducting gap versus half width $a$ within the window $0.5 \le a(nm) \le 2.5$, as obtained from Eqs.(\ref{gapzero}) and (\ref{gapinfty}).
The two selected points marked by an ``asterisk'' represent the equivalence between these two curves. The $\Omega=0$ and the $\Omega=\infty$ curves are equivalent by a scaling $a \rightarrow 2a$.
The  black horizontal line sets the value of the bulk gap.
} \label{delta-a-zero-infty}
\end{center}
\end{figure}
\begin{figure}[hb]
\begin{center}
\includegraphics[width=1.0\linewidth]{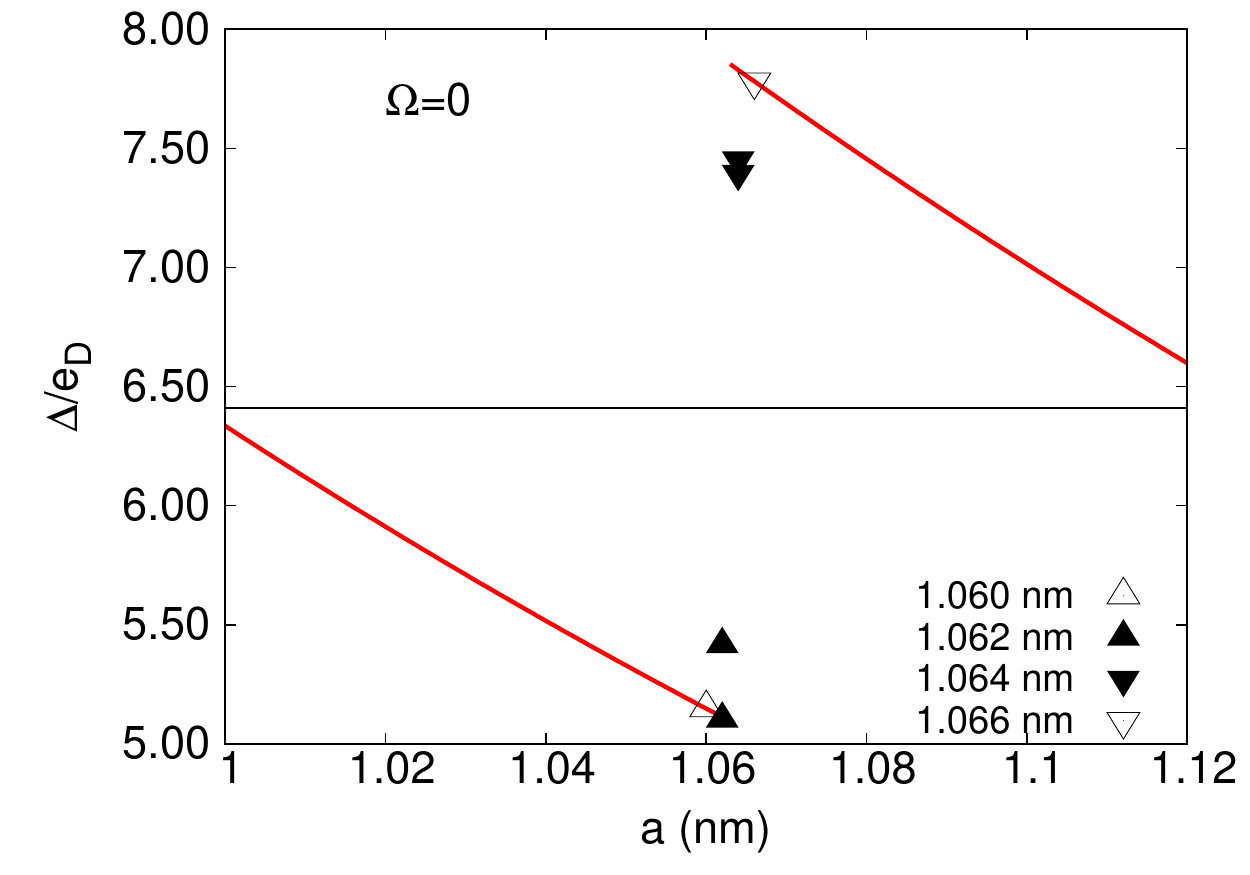}
\caption{(Color on line) The superconducting gaps are shown through a shape resonance windows for the case of no insulating barrier ($\Omega=0$). In this situation the SIS nanofilm becomes a single slab nanofilm of thickness $2a$.
The (red) curve correspond to the superconducting gap obtained from Eq.(\ref{gapzero}) whereas
at the half widths $a=$ 1.060, 1.062, 1.064 and 1.066 nm  the superconducting gaps are depicted by triangles, and obtained from Eq.(\ref{gapeq2}). There is just one single superconducting gap at the extremities of the resonance window ($a=$ 1.060 and 1.066 nm). A second superconducting gap unfolds within the resonance window ($a=$ 1.062 and 1.064 nm) which is characteristic of a shape resonance.
The  black horizontal line sets the value of the bulk gap.
}\label{shape}
\end{center}
\end{figure}
\begin{figure}[hb]
\centering
\begin{center}
\includegraphics[width=0.8\linewidth]{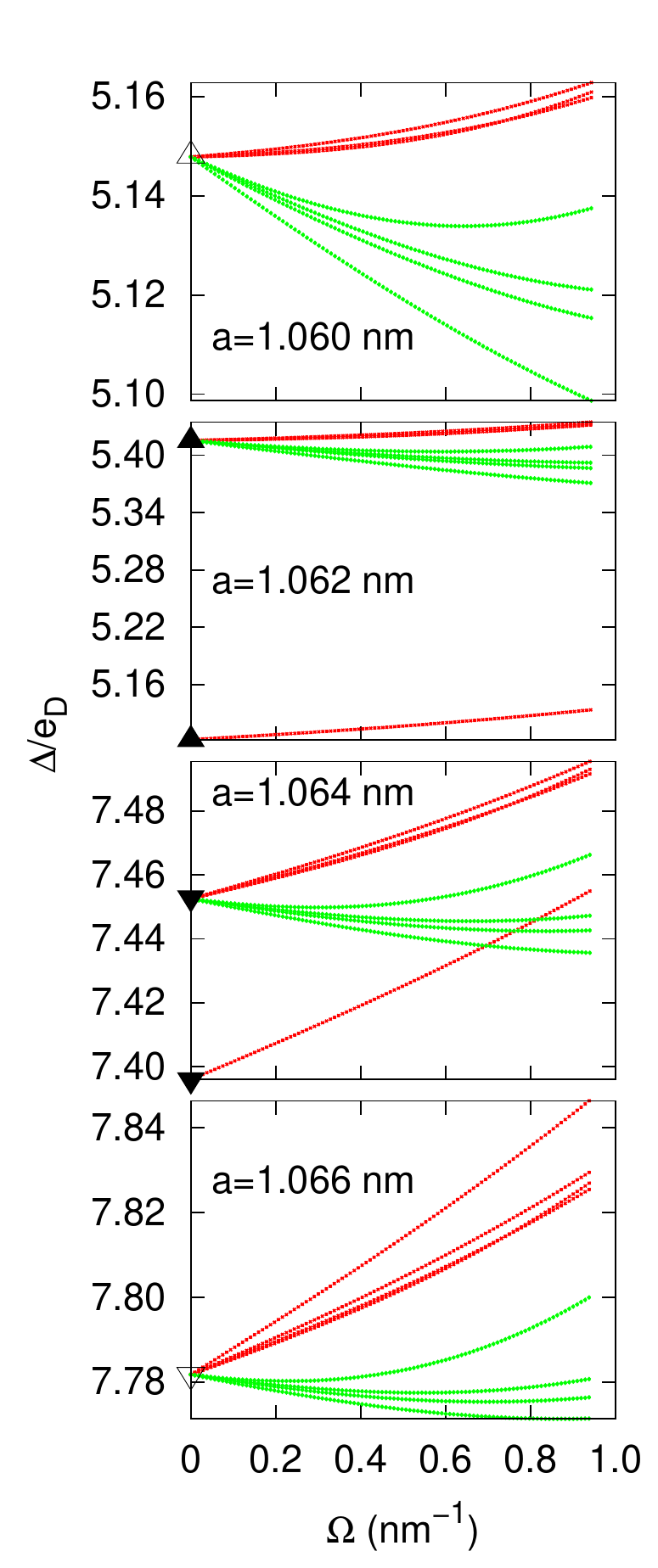}
\caption{(Color on line) The superconducting gaps  of the SIS nanofilm for half widths $a=$ 1.060, 1.062, 1.064 and 1.066 nm versus the insulating barrier strength $\Omega$. The $\Omega=0$ gaps are marked by triangles, also shown in Fig.~\ref{shape}.
Even and odd gaps are presented by gray (green) and dark (red) lines , respectively.
The insulating barrier ($\Omega\neq 0$) lifts the degeneracy among the superconducting gaps.
For $\Omega= 0$ the multigaps, also shown in  Fig.~\ref{shape}, are due to the shape resonance.
}\label{multiple}
\end{center}
\end{figure}
\begin{figure}[hb]
\begin{center}
\includegraphics[width=1.0\linewidth]{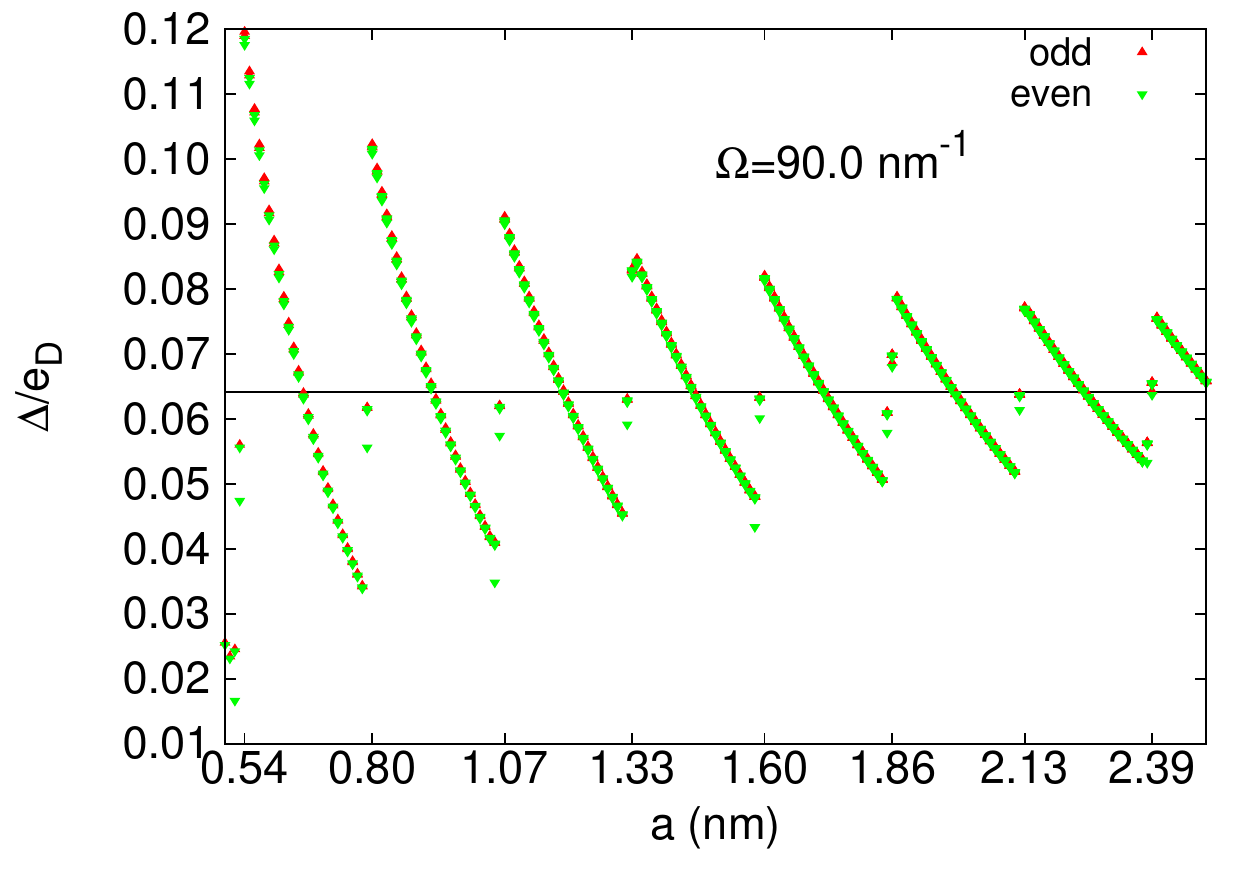}
\caption{(Color on line)
The superconducting gap versus half thickness $a$ within the window $0.5 \le a(nm) \le 2.5$, the same of Eq.(\ref{delta-a-zero-infty}).
This plot shows that  shape resonances are also present in case of a finite  ($\Omega\neq 0$) barrier.
Even and odd gaps are presented by gray (green) and dark (red) lines, respectively.
The x-axis thick marks are set at the top of the individual branches, where is the onset of single particle states with equal number of even and odd states, according to Table~\ref{tab8}.
The  black horizontal line sets the value of the bulk gap.
}\label{delta-a-omega}
\end{center}
\end{figure}
\begin{figure}[hb]
\begin{center}
\includegraphics[width=1.0\linewidth]{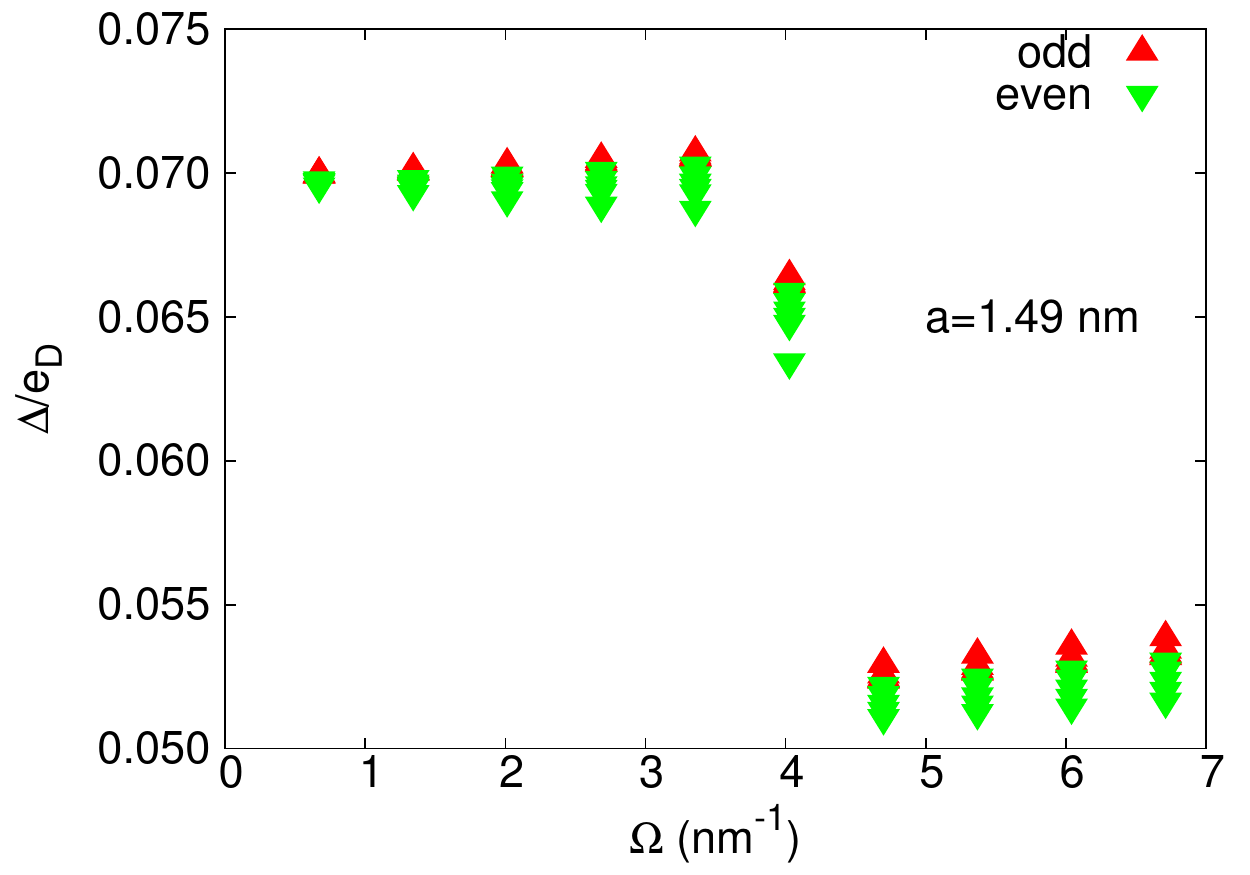}
\caption{(Color on line) An interaction driven resonance is shown here for
a fixed set of ten insulating barrier strengths, given by $a\Omega = 1,2,\ldots 10$. Even and odd gaps are presented by gray (green) and dark (red) lines, respectively.
The single particle energy levels for selected $\Omega$ values associated to this transition are described in Tables~\ref{tab7} and \ref{tab5}.
The multigap structure of the SIS nanofilm is seen for a fixed $\Omega$.
}\label{a_149}
\end{center}
\end{figure}
\begin{figure}[hb]
\begin{center}
\includegraphics[width=1.0\linewidth]{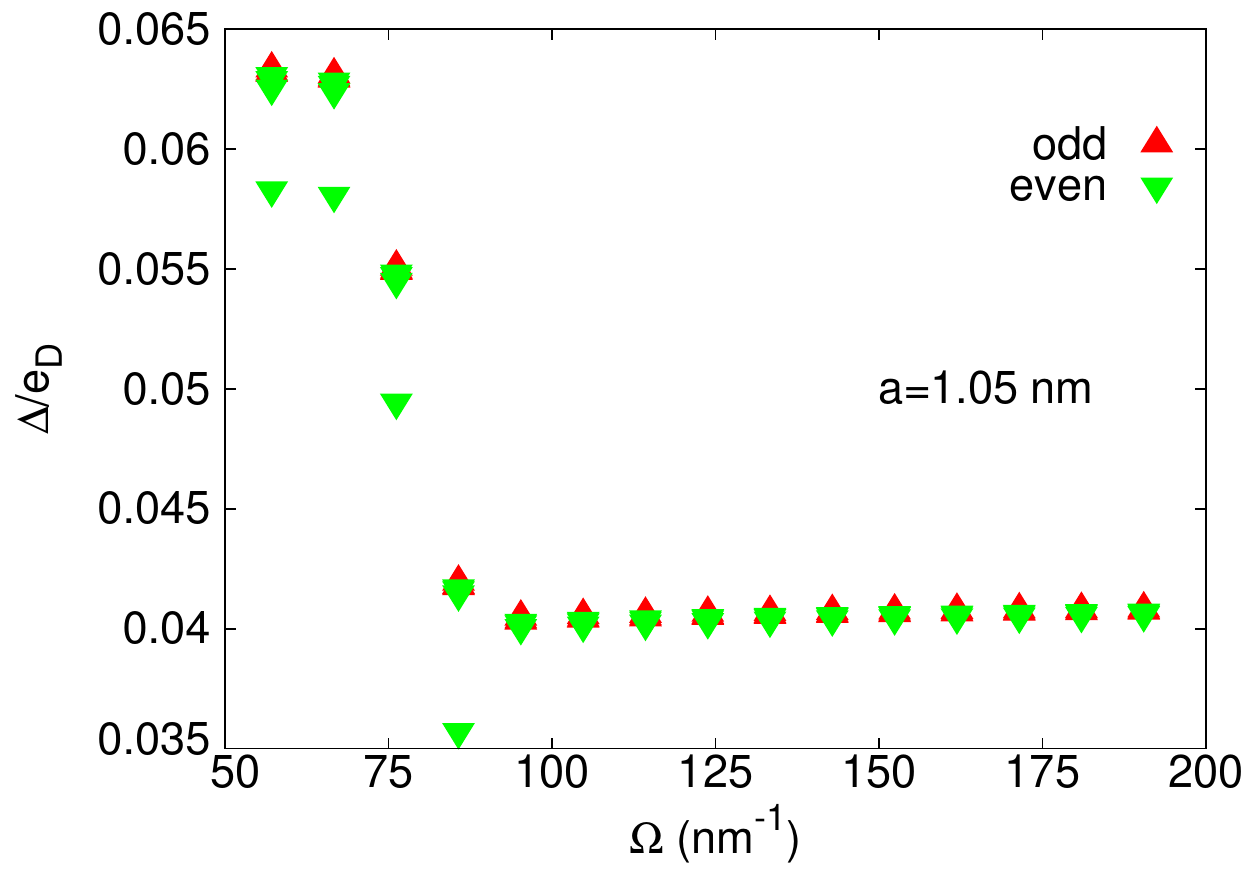}
\caption{(Color on line) An interaction driven resonance is shown here  within the range $50.0<\Omega < 200.0$ for a specific half width. Notice that this range falls below the critical value $\Omega_c$ of Eq.(\ref{omegc}) and Table~\ref{tab2}. Even and odd gaps are presented by gray (green) and dark (red) lines , respectively.
The single particle energy levels for selected $\Omega$ values associated to this transition are described in Table~\ref{tab5}.
The distinct gaps are seen  at fixed $\Omega$ showing the multigap structure of the SIS nanofilm.
}\label{a_105}
\end{center}
\end{figure}

\begin{table}[htb]
\centering
\begin{tabular}{|c||c|c|c|c|c|}
\hline
metal & T$_c$ (K) & $\Delta_{bulk}$ (K) & E$^{3D}_F$ (10$^4$ K) & $e_D$ (K) & k$_F$ (nm$^{-1}$)\\ \hline \hline
{\it Al} & 1.2 & 1.97 & 13.6 & 433 & 17.5\\ \hline
{\bf Nb} & 9.3 & 17.7 &6.18 & 276 & 11.8\\ \hline
{\it Pb} & 7.2 & 15.8 &11.0 & 105 & 15.8 \\ \hline
\end{tabular}
\caption{The Aluminum, Lead and Niobium parameters required for the study of the SIS nanofilm. The critical temperature is given for completeness since our study is restricted to $T=0$. The bulk gap, $\Delta_{bulk}$, is retrieved in the limit of a very thick film from Eq.(\ref{deltai}). The bulk Fermi surface, $E^{3D}_{F}$, sets the chemical potential for $T=0$, as given by Eq.(\ref{mut0}). The Fermi wavelength, $k_F$, enters in Eq.(\ref{deltai}), as shown in Ref.~\onlinecite{thompson63}. The Debye energy, $e_D$, is defined in Eq.(\ref{debye}).}
\label{tab1}
\end{table}
\begin{table}[htb]
\centering
\begin{tabular}{|c||c|c|c|c|c|}
\hline
metal & $\mu/e_D$  & $\epsilon_0/e_D$ & $\Delta_{bulk}/e_D$ & $\lambda$ & $\Omega_c$ (nm$^{-1}$)\\ \hline \hline
{\it Al} & 314.1 & 10.08 & 4.56 $\cdot$ 10$^{-3}$ & 0.1640& 2.21 $\cdot$ 10$^{3}$ \\ \hline
{\bf Nb} & {\bf 223.9} & {\bf 15.81} & {\bf 6.41} $\cdot${\bf 10} $^{{\bf-2}}$ & {\bf 0.2901}& {\bf 2.46}  $\cdot$ {\bf 10}$^{{\bf 2}}$\\ \hline
{\it Pb} & 1048 & 41.56 & 1.51  $\cdot$ 10$^{-1}$ & 0.2889 & 2.75  $\cdot$ 10$^{2}$ \\ \hline
\end{tabular}
\caption{The Aluminum, Lead and Niobium energy parameters required for the study of the SIS nanofilm are shown here in units of the Debye energy, $e_D$, defined in Eq.(\ref{debye}). The coupling $\lambda$ is set so to obtain the bulk gap, $\Delta_{bulk}$, from Eq.(\ref{deltai}) in the limit of a very thick film. The critical barrier strength, $\Omega_c$, is obtained from Eq.(\ref{omegc}) for Aluminum, Niobium and Lead.}
\label{tab2}
\end{table}
\begin{table}[htb]
\centering
\begin{tabular}{|c||c||c|}
\hline $\Omega=0$ & $a=$ 1.060 & $a'=$ 1.066\\
& nm & nm \\
\hline
 symmetry & $\epsilon_j(a)/e_D$ & $\epsilon_j(a')/e_D$  \\
\hline \hline
{\it even} & 3.518 \,& 3.479 \, \\
\hline
{\it even} & 31.66 & 31.31\\
\hline
{\it even} & 87.95 & 86.96\\
\hline
{\it even} & 172.4& 170.5\\
\hline
{\it odd} &  14.07 &  13.91\\
\hline
{\it odd} & 56.29 & 55.68\\
\hline
{\it odd} & 126.7 & 125.2\\
\hline \hline
{\it odd} & & 222.6 \\
\hline
\end{tabular}
\caption{
The single particle energy levels in case of no barrier and at the extremities of the shape resonance region defined by $a$ and $a'$, as described in  Fig.~\ref{shape}.  The number of energy levels changes from 7 to 8, as shown here.
The Debye energy window, defined in Eq.(\ref{debye}) is $\mu/e_D-1=222.9$. $\mu/e_D+1=224.9$, using values of Table~\ref{tab2}.
The Debye energy window lies below and above the highest energy level, for $a$ and  $a'$, respectively, thus leading to a shape resonance.
Figs.~\ref{shape} and \ref{multiple} show the onset of multigaps through this transition.}
\label{tab4}
\end{table}

\begin{table}[htb]
\centering
\begin{tabular}{|c||c||c||c|}
\hline $a$ (nm)  & (even, odd)  &  $\begin{array}{c} 10^{2}\Delta^{max}_e/e_D \\ 10^{2}\Delta^{min}_e/e_D\end{array}$ &
$\begin{array}{c} 10^{2}\Delta^{max}_o/e_D \\ 10^{2}\Delta^{min}_o/e_D\end{array}$ \\
\hline
\hline
0.50 & (1,1)& 2.5249 & 2.5526 \\ \hline
0.52 & (2,1)& $\begin{array}{c} 2.4284\\ 1.6662\end{array}$ &2.4528\\ \hline
0.54 & (2,2)& $\begin{array}{c} 11.8548 \\11.7608 \end{array}$ & $\begin{array}{c} 11.9588\\11.8947\end{array}$ \\ \hline
0.79 & (3,2)& $\begin{array}{c} 6.1575\\5.5620\end{array}$ & $\begin{array}{c}6.1786 \\ 6.1679 \end{array}$ \\ \hline
0.80 & (3,3)& $\begin{array}{c} 10.1640 \\10.0861\end{array}$ &  $\begin{array}{c}10.2309\\10.1757 \end{array}$\\ \hline
1.05 &  (4,3) &$\begin{array}{c} 4.0852\\ 3.4853\end{array}$ & $\begin{array}{c} 4.1063 \\4.0888 \end{array}$\\ \hline
1.07 &  (4,4) & $\begin{array}{c} 9.0633\\9.0016 \end{array}$ & $\begin{array}{c} 9.1129\\9.0680 \end{array}$ \\ \hline
1.32 &  (5,4) & $\begin{array}{c} 6.2923\\5.9186 \end{array}$&$\begin{array}{c} 6.3092\\6.2945 \end{array}$ \\ \hline
1.33  & (5,5) & $\begin{array}{c} 8.2935\\8.1921 \end{array}$ & $\begin{array}{c} 8.2958\\8.2821 \end{array}$ \\ \hline
1.58 &  (6,5) & $\begin{array}{c} 4.7967\\ 4.3375\end{array}$ & $\begin{array}{c} 4.8165\\4.7978 \end{array}$ \\ \hline
1.60 &  (6,6)& $\begin{array}{c} 8.1720 \\8.1273 \end{array}$ & $\begin{array}{c} 8.2070\\8.1733 \end{array}$ \\ \hline
1.85 &  (7,6) & $\begin{array}{c} 6.0919 \\5.7909 \end{array}$ & $\begin{array}{c} 6.1078\\6.0927 \end{array}$ \\ \hline
1.86 &  (7,7) & $\begin{array}{c} 6.9810 \\6.8026 \end{array}$ & $\begin{array}{c} 6.8637\\6.9818 \end{array}$\\ \hline
2.12 &  (8,7) & $\begin{array}{c} 6.3718\\6.1406 \end{array}$ & $\begin{array}{c} 6.3857\\6.3723 \end{array}$\\ \hline
2.13 &  (8,8)& $\begin{array}{c} 7.7009\\7.6606 \end{array}$ & $\begin{array}{c} 7.7232\\7.7015 \end{array}$\\ \hline
2.38 &  (9,8)& $\begin{array}{c} 5.6252\\5.3265 \end{array}$ & $\begin{array}{c} 5.6415\\5.6256 \end{array}$\\ \hline
2.39 &  (9,9)& $\begin{array}{c} 6.5536\\6.3743 \end{array}$ & $\begin{array}{c} 6.4225\\6.5539 \end{array}$\\ \hline
\end{tabular}
\caption{This table contains the half widths where there are changes in the number of even and odd single particle states, referred by the pair of integers $(e,o)$. The barrier strength is $\Omega=$ 90 nm$^{-1}$.
The number of even (odd) single particle states  is the same as the number of distinct even (odd) superconducting gaps whose maximum and minimum values are shown here. These gaps are also seen in Fig.\ref{delta-a-omega}
Notice that the highest superconducting gaps are always odd while the lowest ones are even. }
\label{tab8}
\end{table}
\section{Bogoliubov de Gennes equations and Anderson approximation for SIS nanofilms}\label{BdG}
The energy of the excitations above the superconducting ground state at zero temperature, $E_n$, are obtained from the BdG equations.
\begin{eqnarray} \label{bdg0}
\left[
\begin{array}{cc}
\displaystyle H_0(\textbf{r})
& \displaystyle\Delta(\textbf{r})\\
\displaystyle\Delta^{*}(\textbf{r})
& -\displaystyle H_0(\textbf{r})\\
\end{array}
\right]
\left[
\begin{array}{c}
\displaystyle u_n(\textbf{r})\\
\displaystyle v_n(\textbf{r})\\
\end{array}
\right] = E_n
\left[
\begin{array}{c}
\displaystyle u_n(\textbf{r})\\
\displaystyle v_n(\textbf{r})\\
\end{array}
\right]
\end{eqnarray}
where the single particle Hamiltonian is,
\begin{eqnarray}\label{hamil}
H_0(\textbf{r})= - \frac{\hbar^2}{2m}\nabla^2-\mu +U(\textbf{r}).
\end{eqnarray}
The potential $U(\textbf{r})$ confines particles inside the thin film and may contain an internal structure.
The BdG equations are non-linear in $u_n(\textbf{r})$ and $v_n(\textbf{r})$ since,
\begin{eqnarray} \label{gap0}
\Delta(\textbf{r})=V\sum_n v^*_n(\textbf{r})u_n(\textbf{r}),
\end{eqnarray}
where $V$ is the strength of the pairing interaction ($V>0$ in the following).
The number of particles and the density are given by,
\begin{eqnarray}\label{dens}
 N \equiv \int d\textbf{r} \, n(\textbf{r}), \quad n(\textbf{r})=2\sum_n |v_n(\textbf{r})|^2.
\end{eqnarray}
One seeks normalized wavefunctions,
\begin{eqnarray}
\int d\textbf{r} \, \left [ u^*_n(\textbf{r})u_m(\textbf{r})
+ v^*_n(\textbf{r})v_m(\textbf{r}) \right ] = \delta_{n,m}.
\end{eqnarray}
In case of a single superconducting slab the summation index represents the set of parallel and perpendicular degrees of freedom,
associated to continuous and discrete wave numbers, respectively.

The Anderson approximation relies on the Schroedinger equation,
\begin{eqnarray}\label{schro-eq}
H_0(\textbf{r})\Psi_n(\textbf{r})  = \zeta_n \Psi_n(\textbf{r}),
\end{eqnarray}
whose eigenstates provide the approximated solution to the BdG equations,
\begin{eqnarray} \label{sol-bdg}
&& u_n(\textbf{r})=c_n\Psi_n(\textbf{r}), \nonumber\\
&& v_n(\textbf{r})=d_n\Psi_n(\textbf{r}),
\end{eqnarray}
where the eigenvalues, $\zeta_n$, and normalized eigenvectors,
\begin{eqnarray} \label{norm2}
\int d\textbf{r} \, \psi^*_n(\textbf{r})\Psi_m(\textbf{r})
= \delta^{}_{n,m}.
\end{eqnarray}
The coefficients $c_n$ and $d_n$ are the coherence factors of the BCS-like superconducting state and are given by,
\begin{eqnarray}
&& c_n = \left [\frac{1}{2}\left (1+\frac{\zeta_n}{E_n} \right)\right ]^{1/2}, \label{cd1}\\
&& d_n = \left [\frac{1}{2}\left (1-\frac{\zeta_n}{E_n} \right)\right ]^{1/2}, \, \mbox{and} \label{cd2}\\
&& E_n = \sqrt{\zeta_n^2+\Delta_n^2},
\end{eqnarray}
where
\begin{eqnarray}
\Delta_n\equiv \int d\textbf{r} \, \Delta(\textbf{r})|\Psi_n(\textbf{r})|^2.
\end{eqnarray}
Therefore the excitation energies $E_n$ are given in terms of the superconducting gaps $\Delta_n$ that must obey self consistent equations, equivalent to Eq.(\ref{gap0}).
\begin{eqnarray}
\Delta_n = \sum_m V_{n\,m}\frac{\Delta_m}{2\sqrt{\zeta_m^2+\Delta_m^2}}.
\end{eqnarray}
The pairing interaction is given by the matrix elements,
\begin{eqnarray}\label{vnm}
V_{n\,m}=V\int d \textbf{r}\; |\Psi_n(\textbf{r})|^2 |\Psi_m(\textbf{r})|^2.
\end{eqnarray}
The Anderson solution only solves an approximate version of the BdG equations,
as can be deduced by introducing the proposed solution,  Eq.(\ref{sol-bdg}), into Eq.(\ref{bdg0}),
\begin{eqnarray}
&& \zeta_n c_n \Psi_n(\textbf{r})+\Delta_n(\textbf{r})d_n \Psi_n(\textbf{r})=E_n c_n \Psi_n(\textbf{r}) \nonumber \\
&& \Delta^{*}(\textbf{r})c_n\Psi_n(\textbf{r})-\zeta_n d_n \Psi_n(\textbf{r})=E_n d_n \Psi_n(\textbf{r}).
\end{eqnarray}
Only the integrated version of the above equations, multiplied by $\Psi^{*}_n(\textbf{r})$, is solved. Using the normalization condition of Eq.(\ref{norm2}), one obtains the Anderson approximation equations, which corresponds to the set of equations below, which is position independent.
\begin{eqnarray} \label{bdg1}
\left[
\begin{array}{cc}
\zeta_n &  \Delta_n\\
\Delta_n & - \zeta_n\\
\end{array}
\right]
\left[
\begin{array}{c}
\displaystyle c_n\\
\displaystyle d_n\\
\end{array}
\right] = E_n
\left[
\begin{array}{c}
\displaystyle c_n\\
\displaystyle d_n\\
\end{array}
\right].
\end{eqnarray}
The solution of the above equations are given by Eqs.(\ref{cd1}) and (\ref{cd2}). They are the Anderson approximation to the  solution of the full BdG equations.

The BdG equations applied to the SIS nanofilm must take into account the perpendicular and parallel degrees of freedom of electron motion, such that
the single particle states, defined by Eq.(\ref{schro-eq}), have the following properties:
\begin{eqnarray}
&& n \equiv \left ( \vec k_{\parallel},k_j\right) \label{indice} \\
&& \zeta_{n}=\epsilon_{\parallel}+\epsilon_{j}-\mu, \quad
\epsilon_{\parallel}=\frac{\hbar^2 \vec k_{\parallel}^2}{2m}, \quad \epsilon_{j}=\frac{\hbar^2 k_j^2}{2m} \label{sumener}\\
&&\psi_n(\vec x_{\parallel},x) = \frac{1}{\sqrt{A}}e^{i\vec k_{\parallel}\cdot \vec x}\psi_j(x).
\end{eqnarray}
The nanofilm surface area is $A$ and the wave function component $\psi_{k_j}(x)$ satisfies the uni-dimensional Schroedinger equation,
\begin{eqnarray}\label{schro-delta}
-\frac{\hbar^2}{2m}\frac{d^2\psi_{k_j}}{dx^2}+U(x)\psi_{k_j}=\epsilon_{k_j}\psi_{k_j}.
\end{eqnarray}
All $k_j$ states, $j$ being an integer, are bounded and for simplicity also labeled by $k$ or just by $j$. The wavefunctions are assumed normalized.
\begin{eqnarray}\label{norm1}
\int^a_{-a}dx \, \psi^*_{j}(x)\psi^{}_{j'}(x) = \delta^{}_{j\,j'}.
\end{eqnarray}
Thus the general Eqs.(\ref{schro-eq}) and (\ref{norm2}) are reduced to the perpendicular ones, Eqs.(\ref{schro-delta}) and (\ref{norm1}), respectively.
The dimensionless interaction matrix component in Eq.(\ref{vnm}), associated to these wave functions defined below.
\begin{eqnarray}\label{vij}
V_{i\,j} \equiv 2a\int^a_{-a}dx \, |\psi_i(x)|^2 \, |\psi_{j}(x)|^2,
\end{eqnarray}
and $2a$ is the total width of the SIS nanofilm.

\section{Description of the insulating slab}\label{sis3l}
The total thickness of the SIS nanofilm and of the insulating barrier are $2a$ and $2\varepsilon$, respectively, as depicted in Fig.~\ref{sandwich}.  The insulating slab is a potential barrier of height $V_0$ that acts in the single particle channel of the BdG equations, namely, described by the Hamiltonian of Eq.(\ref{hamil}), through the potential that confines particles inside the SIS nanofilm.
\begin{eqnarray} \label{finitebarrier}
U(x) = \left\{
\begin{array}{rcl}
\infty,& \mbox{if} & |x| \ge a\\
0, & \mbox{if} & \varepsilon \le |x|<a\\
V_b(x), & \mbox{if} & |x| < \varepsilon
\end{array}
\right.
\end{eqnarray}
Fig.~\ref{pot-well} depicts this rectangular barrier of height $V_b(x)=V_0$ and two single particle energies, representative of states that lie above and below the barrier. Many properties of the single particle states in presence of the potential barrier can be understood without solving the Schroedinger equation (Eq.(\ref{schro-delta})).
This is because the parity (even-odd) symmetry is broken by the barrier which splits the single particle states  into states that are strongly affected by the presence of the barrier and others that are not.
The odd states are very little sensitive to the barrier because their wavefunction vanishes inside the barrier.
Oppositely, changes in the strength of the barrier severely affect the even states since they dos not vanish inside it.
In this paper we work within the approximation that the odd levels are absolutely not affected by the barrier, which means a very thin insulating slab. In this case the potential barrier of the SIS nanofilm can be approximated by a delta function potential~\cite{flugge47},
\begin{eqnarray}
V_b(x)=\frac{\hbar^2}{m}\Omega \, \delta \left( x \right)=V_0\delta\left( \frac{x}{2\varepsilon}\right).
\end{eqnarray}
We find convenient to describe  the insulating barrier through a characteristic wave number, $\Omega$, related to $V_0$ and $\varepsilon$  by the assumption of equivalence of the rectangular barrier area.
\begin{eqnarray} \label{omegadef}
\frac{\hbar^2}{m}\Omega \equiv \int^{\; a}_{-a} V_b(x) \, dx = V_0 \, 2\varepsilon \rightarrow \Omega = V_0 \frac{2m\varepsilon}{\hbar^2}
\end{eqnarray}
The characteristic energy of the discrete levels, caused by the perpendicular  confinement is
\begin{eqnarray}\label{eps0}
\epsilon_0\equiv \frac{\hbar^2}{2m}\left (\frac{\pi}{x_0}\right )^2=0.38\; \mbox{meV}, \, x_0 \equiv 1.0 \; \mbox{nm}.
\end{eqnarray}
Then one obtains that,
\begin{eqnarray}
V_0 = \epsilon_0 \left (\frac{x_0}{\pi^2\varepsilon}\right) \left(x_0\Omega \right ).
\end{eqnarray}
To get a rough estimate for it, consider $\varepsilon \sim  x_0$, then $V_0 \sim 0.038 \,\Omega$  meV, where $\Omega$ must be expressed in nm$^{-1}$.

The delta function potential description provides a simplified model that retains key features of the barrier.
There are single particle states above and  below the barrier, such as shown in Fig.~\ref{pot-well}.
The perpendicular discrete levels remain parabolic for any $\Omega$, $\epsilon_j= \hbar^2k_j(\Omega)^2/2m$.
We briefly summarize some key properties of the even and odd energy levels  within the delta function description of the barrier.
Even and odd single energy levels alternate in sequence, as shown in Figs.~\ref{even-odd-vs-omega} and ~\ref{levels-vs-omega}.
The even levels have their  wavenumber  determined by
\begin{eqnarray} \label{kcotk}
ka \cdot \cot ka =- \Omega a,
\end{eqnarray}
where for a given $\Omega$ there are several $k_j$ solutions. They are affected by the presence of the insulating barrier, $\epsilon_{j}=\epsilon_{j}(\Omega)$, as previously described. The odd  wavenumbers are simply given by
\begin{eqnarray} \label{ksym}
k_j=j\pi/a, \quad j=1,2,3,\ldots.
\end{eqnarray}
The physical interpretation of $\Omega$ is a crossover wavenumber.
It sets the regime beyond which the barrier is no longer felt, thus in agreement with Fig.~\ref{pot-well}.
To see this notice that Eq.(\ref{kcotk}) can be written as $ka \cdot \cos ka +\Omega a \cdot \sin ka=0$.
For $|k|>>\Omega $ the eigenvalue equation simply becomes $\cos ka \approx 0$ since the trigonometric functions are bounded.
Thus even modes become $k_j\approx (2j+1)\pi/a$, $j=0,1,2 \ldots$. They do not feel the barrier and can be considered as lying above it.
For $|k|<<\Omega $ the eigenvalue equation must be solved and in this case the even modes feel the barrier. In summary $\Omega$ is the critical wavenumber associated to the barrier height.\\

To reach understanding of the effects of the potential barrier into the single particle energy levels, we consider  three special values of the potential barrier strength, namely, $\Omega=0$, finite, and $\infty$.\\

Fig.~\ref{even-odd-vs-omega}  shows the single particle ground and first excited states, which are even and odd states, respectively.
While the ground state is deformed by the potential barrier, the first excited state remains unaffected.
They are quite distinct for $\Omega=0$, but as $\Omega\rightarrow\infty$ the ground state becomes equal to the $\Omega=0$ ground state, although doubled due to the presence of the infinite barrier.
Moreover the ground and first excited states become equivalent in this limit since even and odd levels coincide apart from a negative phase difference in one of the half-widths.\\

Fig.~\ref{levels-vs-omega} gives a pictorial view of the discrete levels, as a function of $ka$, in these three situations. It also shows that from $\Omega=0$ to  $\Omega \rightarrow \infty$ even and odd levels approach each other and confirms the degeneracy seen in
Fig.~\ref{even-odd-vs-omega} for the fundamental and first excited states here extended for all other even and odd excited levels.
As even and odd levels alternate in energy at the limit $\Omega = \infty$ they become degenerate in energy.\\

Fig.~\ref{chemical-potential-bands} describes the  multiple Fermi surfaces in the nanofilm.
The first panel shows the several parabolic bands and their intersection with the Debye energy window defined by Eq.(\ref{debye}).
It takes into account the alternate presence of even and odd levels, which is a general property of the SIS nanofilm.
The second panel shows the result of the intersection of the parabolic bands with the chemical potential, which at zero temperature is fixed and equal to the three-dimensional Fermi energy of the bulk system,
\begin{eqnarray}\label{mut0}
\mu \equiv E^{3D}_F.
\end{eqnarray}
At zero temperature, the multiple two-dimensional Fermi surfaces are defined by,
\begin{eqnarray} \label{fermi2d}
\epsilon_{Fj}^{2D}= \mu -\epsilon_{j} >0.
\end{eqnarray}
Thus even and odd Fermi surfaces succeed in sequence of increasing $j$ until a  maximum $j_{max}$ is reached, and beyond which it is no longer possible to have a positive $\epsilon_{Fj}^{2D}$.
The crossing of the Debye energy window,
\begin{eqnarray}\label{debye}
\delta \mu_D \equiv (\mu-e_D, \mu+e_D),\; \mbox{where} \; e_D \equiv \hbar \omega_D,
\end{eqnarray}
centered at $\mu$, with these multiple bands, defined by $\epsilon_{j}$, is shown in Fig.~\ref{chemical-potential-bands}.
As well-known, pairing occurs only within the Debye energy window around  each of these two-dimensional Fermi surfaces.

An interesting feature of the SIS nanofilm is the equivalence of its two extreme limits concerning the barrier strengt:
$\Omega \rightarrow 0$ and also $\Omega \rightarrow\infty$ should yield the same physics  since the single slab nanofilm is retrieved in both limits.
For $\Omega \rightarrow\infty$ the SIS nanofilm becomes two independent disconnected single slab nanofilms with half the width of the $\Omega \rightarrow 0$ case.
However close to a critical value of the barrier, $\Omega_c$ the present BdG formalism and its Anderson approximation break down. This is because the splitting between consecutive even and odd single particle energy levels becomes smaller than the superconducting gap.
The present formalism does not contemplate the possibility of pairing between particles in distinct energy levels (cross pairing) and only considers pairing between particles in the same energy level, with the possibility of Cooper pair transfer between different subbands.

\section{Multiband and multigap structure of SIS nanofilms}\label{gapeqs}
The insulating barrier lifts the degeneracy of even and odd single particle states, because it acts differently on them, rendering the superconducting gap in SIS nanofilm very distinct from the one in single slab nanofilm. Therefore the SIS nanofilm can be regarded as a tunable system to study the interplay between parity symmetry and superconductivity. The SIS and the single slab nanofilms are both multiband systems but with very distinct multigap structure, as shown here. The single slab features multigaps only within the shape resonance region whereas the SIS nanofilm is multigapped independently of any resonance condition.

We assume here that for all the 2D Fermi surface the Debye energy window is the same, similarly to the Thompson-Blatt model for the single slab nanofilm~\cite{thompson63}. The effective pairing interaction in the BCS like approximation exists only within the Debye energy window. According to Eq.(\ref{indice}) one obtains that,
\begin{eqnarray}
V_{n\,m} = V_{i\,j}\theta\left (e_D-|\zeta_{\vec k_{\parallel},i}| \right )\cdot
\theta\left (e_D-|\zeta_{\vec k'_{\parallel},j}| \right ),
\end{eqnarray}
where the matrix elements were defined in Eqs.(\ref{vnm}) and (\ref{vij}), and $\zeta_{\vec k_{\parallel},j}$ is defined in Eq.(\ref{sumener}). The Heavyside function, $\theta(x)$, has been employed: $\theta(x)=1$ for $x>0$ and $\theta(x)=0$ for $x<0$. The superconducting gap is homogeneous within each Debye energy window,
\begin{eqnarray}
\Delta_n \equiv \Delta_j\left ( \vec k_{\parallel}\right )=\Delta_j\theta\left (e_D-|\zeta_{\vec k_{\parallel},j}| \right ),
\end{eqnarray}
There is a maximum parallel energy $\epsilon^{max}_{\parallel}=E_{2j}$ and a minimum one, $\epsilon^{min}_{\parallel}=E_{1j}$, which defines the parallel energy window, $\delta E_{\parallel}\equiv E_{2j}-E_{1j}$.
\begin{eqnarray}
E_{2j}&=&\left ( \mu+e_D  -\epsilon_{j}\right )\theta\left [ \mu+e_D  -\epsilon_{j}\right ], \label{e2j}\\
E_{1j}&=&\left ( \mu -e_D  -\epsilon_{j}\right )\theta\left [ \mu -e_D  -\epsilon_{j}\right ]. \label{e1j}
\end{eqnarray}
Let us analyze this parallel energy window according to the three possible positions of the Debye energy window:
\begin{enumerate}
\item {\it $\delta\mu_D $  is above the bottom of the $\epsilon_j$ band.} Then $E_{2j}= \mu+e_D  -\epsilon_{j}$ and $E_{1j}= \mu-e_D  -\epsilon_{j}$ and
    $\delta E_{\parallel}=2e_D $.
\item {\it $\mu$ touches the bottom of the band, namely, $\mu=\epsilon_{j}$}. Then
    $E_{2j}= \mu+e_D  -\epsilon_{j}$ and $E_{1j}= 0$, thus
    $\delta E_{\parallel}=e_D $.
\item {\it $\mu+e_D$ is below $\epsilon_j$}. Then   $E_{2j}= 0$ and $E_{1j}= 0$, and consequently $\delta E_{\parallel}=0$.
\end{enumerate}
In this paper we take the assumption that $\mu >> e_D$. This renders the window  $\delta \mu_D $  very narrow and consequently, also the (shape) resonance window small. We also take that the parallel electronic density of states is constant, and given by the standard definition,
\begin{eqnarray}
N_{2D}/A= \frac{m}{2\pi\hbar^2}=\frac{\pi}{4 x_0^2\epsilon_0}.
\end{eqnarray}
Therefore the summation over single particle states is really given by $\sum_n (\ldots)=\sum_{\vec k_{\parallel},\, j}(\ldots) =  N_{2D}\int^{E_{2j}}_{E_{1j}}d\epsilon_{\parallel}\sum_i(\ldots)$.
The equation that determines the superconducting gaps is obtained by integrating over the parallel momentum:
\begin{eqnarray}
& \Delta_i = \frac{\pi \lambda}{k_F 2a}\sum_j V_{i\,j} F(\Delta_j)\Delta_j , \, \mbox{where} \label{deltai} \\
& F(\Delta_j) = \frac{1}{2}
\left \{\sinh^{-1} \left[ \frac{E_{2j}+\epsilon_j-\mu}{\Delta_j} \right]
 -\sinh^{-1} \left[ \frac{E_{1j}+\epsilon_j-\mu}{\Delta_j} \right]
\right\} \nonumber \\ \label{Ffunc}
\end{eqnarray}
where the interaction matrix has been defined in Eq.(\ref{vij}).

We have expressed the coupling $V$, introduced in Eq.(\ref{vnm}), in terms of $\lambda$, previously defined in Table~\ref{tab2}:
\begin{eqnarray}\label{vlamb}
\lambda=\frac{ k_F N_{2D} V}{\pi A }
\end{eqnarray}
where $k_F$ is the three-dimensional Fermi wavenumber.
Although we do not impose a fixed total number of particles, for the sake of completeness, we show that Eq.(\ref{dens}) becomes,
\begin{eqnarray}
&N=N_{2D}\sum_j \Big \{\left( E_{2j}+E_{1j}\right) - \nonumber  \\
& \left [\sqrt{\left(E_{2j}+\epsilon_j-\mu \right)^2+\Delta_j^2}-\sqrt{\left(E_{1j}+\epsilon_j-\mu \right)^2+\Delta_j^2} \right ]\Big \} \nonumber \\
\end{eqnarray}
Next the special limits of no barrier ($\Omega=0$) and of an infinitely strong barrier ($\Omega=\infty$) are discussed.
They do not yield any novel result, as they reduce to the single slab nanofilm, well described by  Thompson and Blatt~\cite{thompson63}.
Nevertheless they provide an opportunity to understand properties of the SIS nanofilm and for this reason
we obtain the solutions of  Eq.(\ref{deltai}) in these two limits.
The single particle states are obtained in the appendices \ref{omega0} and  \ref{omegai} from our general approach.\\

Fig.~\ref{pot-zero-infty} depicts for  $\Omega=0$, the single particle ground and the first excited states, which are even and odd states, respectively.
This is the single slab nanofilm limit and the  interaction matrix (Eq.(\ref{vij})) becomes $V_{i\,j}=1+\delta_{i\,j}/2$, as follows from our general expression of Eq.(\ref{vkkp}), shown in appendix \ref{matrix}.
The single particle energies are given by Eqs.(\ref{ejzeroeven}) and (\ref{ejzeroodd}) for even and odd levels, respectively.
Assume the chemical potential out of a resonance condition, namely the first among the three possibilities previously listed.
Away from a shape resonance window Eq.(\ref{deltai}) becomes,
\begin{eqnarray}\label{gapeq2}
\Delta_i = \frac{\pi \lambda}{k_F 2a}\sum_{j=1}^{j_{max}} \left ( 1+\frac{1}{2}\delta_{i\,j}\right) \sinh^{-1} \left (\frac{e_D}{\Delta_j} \right)\Delta_j.
\end{eqnarray}
There is a maximum quantum level,
\begin{eqnarray}\label{jmax1}
j_{max}=floor \left( \sqrt{\frac{\mu}{\epsilon_0}} \frac{2a}{x_0}\right ),
\end{eqnarray}
where ``{\it floor}" stands for the lowest integer.
This integer is obtained from Eq.(\ref{e2j}) and defines the number of activated bands.
When all superconducting gaps are equal, $\Delta_i=\Delta$, one can easily solve Eq.(\ref{gapeq2}) and find its value,
\begin{eqnarray}\label{gapzero}
\frac{\Delta}{e_D}=\frac{1}{\sinh \left [ \frac{k_F 2 a}{\pi \lambda \left (j_{max}+\frac{1}{2} \right )}\right ]}
\end{eqnarray}
Indeed this equal superconducting gap solution is found in Thompson and Blatt's original work~\cite{thompson63}.
We also confirm that all superconducting gaps are equal away from resonance from our numerical solution of Eq.(\ref{deltai}), which is solved iteratively.\\

For $\Omega=\infty$ the single particle states are given by Eq.(\ref{ejinfty}) for both even and odd levels. The analysis of this case is instructive because the two single slab nanofilms limit must be recovered.

Fig.~\ref{pot-zero-infty}  depicts the even and odd single particle wave functions for $\Omega=\infty$.
Even and odd single particle energies are equal in this limit.
The wave functions coincide apart from a phase difference in one of the sides of the barrier (negative sign).
This negative sign is irrelevant for the matrix elements of Eq.(\ref{vnm}) since they only take into account the modulus of the wave function.
According to the  general treatment of the matrix elements done in appendix \ref{matrix}, even and odd terms are all equal: $V^{e,e'}_{i\,j}= V^{o,o'}_{i\,j}=V^{e,o'}_{i\,j}=V^{o,e'}_{i\,j}=1+\delta_{i\,j}/2$.
Consequently the superconducting gaps can be considered equal away from the shape resonances. In case that $\mu >> e_D$ the shape resonances are very narrow. Then  $\Delta^{e}_i=\Delta^{o}_i$ and one obtains a single equation to describe even and odd gaps,
\begin{eqnarray}\label{gapeq3}
& \Delta^{e,o}_i = \frac{\pi \lambda}{k_F 2a}\sum_j \left ( 1+\frac{1}{2}\delta_{i\,j}\right)\cdot \nonumber \\
& \left \{ \sinh^{-1} \left (\frac{e_D}{\Delta^{e,o}_j} \right)\Delta^{e,o}_j
+ \sinh^{-1} \left (\frac{e_D}{\Delta^{o,e}_j} \right)\Delta^{o,e}_j
\right \}.
\end{eqnarray}
Recall that we are considering that  $\mu >> e_D$ and the out of resonance condition, the first one  among the three possibilities previously discussed for the chemical potential. Then we obtain from Eq.(\ref{deltai}) that,
\begin{eqnarray}\label{gapeq4}
\Delta_i = \frac{\pi \lambda}{k_F a}\sum_{j=1}^{j'_{max}} \left ( 1+\frac{1}{2}\delta_{i\,j}\right) \sinh^{-1} \left (\frac{e_D}{\Delta_j} \right) \Delta_j.
\end{eqnarray}
The maximum quantum level is easily obtained from Eq.(\ref{e2j}):
\begin{eqnarray}\label{jmax2}
j'_{max}=floor \left( \sqrt{\frac{\mu}{\epsilon_0}} \frac{a}{x_0}\right ).
\end{eqnarray}
Notice that $j'_{max}$ also follows from Eq.(\ref{jmax1}) under the replacement $2a \rightarrow a$.
Numerical analysis of Eq.(\ref{gapeq4}) shows that all gaps are equal away from the shape resonance, $\Delta^o_i=\Delta^e_i=\Delta$.
Under this assumption the superconducting gap can be easily obtained,
\begin{eqnarray}\label{gapinfty}
\frac{\Delta}{e_D}=\frac{1}{\sinh \left [ \frac{k_F  a}{\pi \lambda \left (j'_{max}+\frac{1}{2} \right )}\right ]}.
\end{eqnarray}

In summary the infinite barrier SIS nanofilm ($\Omega=\infty$) is equivalent to the single slab nanofilm ($\Omega=0$),
because Eq.(\ref{gapeq4}) is equivalent to Eq.(\ref{gapeq2}) under the replacement $2a \rightarrow a$.
This replacement $2a \rightarrow a$ also renders Eq.(\ref{gapinfty}) equivalent to Eq.(\ref{gapzero}).\\

Next we consider the breaking of the present BdG approach due to the onset of cross pairing. The energy splitting between even and odd states, associated to the same discrete state, in the limit $a\Omega >> 1$, is

\begin{eqnarray}\label{delinfty}
\delta \epsilon_j \equiv \epsilon^o_j -\epsilon^e_j= j^2\left ( \frac{\pi}{a}\right)^2\frac{\hbar^2}{2m}\frac{1}{1+\Omega a}.
\end{eqnarray}
according to Eqs.(\ref{ejinfty}) and (\ref{kainfty}).
Essentially cross pairing must be considered in case the above single particle energy splitting becomes comparable to the bulk gap. We define the critical strength $\Omega_c$ at $\delta \epsilon_j=\Delta_{bulk}$.
The smallest splitting is in the first level, $j=1$.
As an example consider the case of a SIS nanofilm with total thickness 2.0 nm, that is, $a=x_0$ (see Eq.(\ref{eps0})).
Then the critical strength of the barrier is given by,
\begin{eqnarray}\label{omegc}
x_0\Omega_c=\frac{\epsilon_0}{\Delta_{bulk}}-1,
\end{eqnarray}
where $\epsilon_0$ is defined in Eq.(\ref{eps0}).
Recall that the discreteness along the perpendicular direction is based on $\epsilon_0>>\Delta_{bulk}$.

Notice that there are two important electronic energy differences, namely, between even and odd levels that form a pair, and  between two consecutive pairs. Even in case the even to odd splitting within a pair falls shorter  than the bulk gap, the splitting between the two pairs consecutive pairs remains larger than the bulk gap, and so, justify the discrete treatment in the perpendicular direction.
For instance,
\begin{eqnarray}
\epsilon^o_{j+1}-\epsilon^o_{j}=(2j+1)\left ( \frac{\pi}{a}\right)^2\frac{\hbar^2}{2m}
\end{eqnarray}
Again taking the example of
$j=1$ and $a=x_0$, gives that $\epsilon^o_{j+1}-\epsilon^o_{j}=3\epsilon_0>> \Delta_{bulk}$.

\section{Physical parameters and single particle states}\label{parameters}
We take the parameters of Niobium described in Tables~\ref{tab1} and \ref{tab2} as a basis for the discussion about the multigap and the interaction driven resonance in  SIS nanofilms.
We also list the physical parameters of Aluminum and Lead, although the present study does not address nanofilms made of these metals.

The minimum energy needed to break a Cooper pair into two independent electrons is $2\Delta$. According to the BCS theory the superconducting gap is related to the critical temperature by $2\Delta/k_BT_c \approx 3.5$ and
Aluminum, Lead and Niobium satisfy approximately this relation~\cite{kittel76,ashcroft76,rohlf94} as seen in Table~\ref{tab1}, which also contains other parameters needed for the study of the SIS nanofilm, such as  $\Delta_{bulk}$, E$^{3D}_F$, $e_D$, and the Fermi wavenumber, $k_F$.

Table~\ref{tab2} shows the required parameters in units of the Debye energy. The ratio $\Delta_{bulk}/e_D = 6.41 \cdot 10^{-2}$ is extensively used in our considerations. Table~\ref{tab2} also contains for the three metals the critical strength of the barrier, defined by Eq.(\ref{omegc}).
For $a=x_0$ (see Eq.(\ref{eps0})) the ratio $\mu/\epsilon_0 \sim 10$, thus a few discrete energy levels fall below the chemical potential, which means that a few of the two-dimensional Fermi surfaces can develop a superconducting gap.
It also holds that the Debye energy is much smaller than the typical quantized energy across the film, $\epsilon_0/e_D=15.81$.
The chemical potential is significantly larger than the Debye energy for the three metals, $\mu >> e_D$, which justifies the analysis done in  section \ref{gapeqs}.
The dimensionless parameter $\lambda$, defined in Eq.(\ref{vlamb}), gives the strength of the pairing interaction and is determined from the the requirement that the nanofilm gaps approach the bulk gap in the limit of increasing thickness.

Fig.~\ref{delta-a-zero-infty} shows the gap versus the half width in the limits $\Omega=0$ and $\Omega=\infty$, as obtained from the single gap solutions of Eqs.(\ref{gapzero}) and (\ref{gapinfty}).
To help visualize the equivalence between the $\Omega=0$ and $\Omega=\infty$ limits, the points $a=$1.063 nm and $a=$ 2.126 nm are encircled.
They yield the same gap value $\Delta/e_D=7.05 \cdot 10^{-2}$.

We use the parameters of Niobium previously assigned to describe the SIS nanofilm.
The shape resonances correspond to the discontinuities where the gap jumps from below to above its bulk value. The bulk gap value is represented there by a black horizontal line.
Indeed Eqs.(\ref{gapzero}) and (\ref{gapinfty}) do not give a good description near to the shape resonances, where in fact a multigap structure arises. There the gap must be directly obtained from Eqs.(\ref{deltai}) and (\ref{Ffunc}). Recall that Eqs.(\ref{gapzero}) and (\ref{gapinfty}) are approximate solutions of Eqs.(\ref{deltai}) and (\ref{Ffunc}) only valid away from resonances.

\section{Multigap features of the SIS nanofilm  at a shape resonance}
In this section a Nb SIS nanofilm is studied with half width in the range $a=$ 1.060 to 1.066 nm and a very weak insulating barrier ($a\Omega << 1$).
Although the studied window is too small to be physical, the present analysis brings understanding about the fundamental differences between the multigap properties of the SIS and the single slab superconducting nanofilms.
Essentially we show here that the SIS nanofilm is intrinsically multigapped whereas the single slab is not.
Through the shape resonance window one observes the onset and disappearance of multiple gaps in the single slab nanofilm while the SIS nanofilm always remains multigapped.\\

Fig.~\ref{shape} shows the gaps in case of no barrier ($\Omega=0$), as obtained through  Eqs.(\ref{gapzero}) and (\ref{deltai}).
The first equation gives the continuous (red) lines while the second one is used for  specific points in the resonance region, namely $a=$ 1.060, 1.062, 1.064 and 1.066 nm.
Recall that Eq.(\ref{gapzero}) is obtained under the assumption of just one single gap and away from resonance. Eq.(\ref{deltai}) provides a self consistent numerical method to numerically determinate the gaps.
Away from resonance they give the same results which proves the righteousness of the gap solution of  Eq.(\ref{gapzero}) for $\Omega=0$.
Indeed the numerical solution of Eq.(\ref{deltai}) shows that the cases $a=$ 1.060 nm and $a=$ 1.066 nm are the extremities of the shape resonance, as  just a single gap is obtained, as shown in Fig.~\ref{shape}.
For the intermediate widths, $a=$ 1.062 nm and 1.064 nm, Eq.(\ref{deltai}) gives two gaps corresponding to a change in the number of single particle states below the chemical potential. Thus  these two half widths fall in the shape resonance window, since a second gap appears, which corresponds to the entrance of a new band into the Debye energy window.
The two gaps at $a=$ 1.062 and 1.064 nm cannot be described by Eq.(\ref{gapzero}) since this formula does not account for  different gaps.
Table~\ref{tab4}  shows the single particle energy levels at the extremities of the studied interval and the onset of a new single particle energy state into the Debye energy window.
The number of even levels remain unchanged while the odd ones are changed by one.
Recall that according to Table~\ref{tab2}, $\mu/e_D=223.9$.

Next consider the situation of a weak potential  barrier, as shown in Fig.~\ref{multiple}.
The four half widths $a=$ 1.060, 1.062, 1.064 and 1.066 nm are considered here in case that $\Omega \neq 0$.
They were considered before in Fig.~\ref{shape} in case that $\Omega = 0$.
Some very distinct features are noticeable in the comparison of the extremities ($a=$ 1.060 nm, 1.066 nm) with the resonating widths  ($a=$ 1.062 nm, 1.064 nm).
At the extremities, the splitting of the gap is purely due to the barrier since for $\Omega = 0$ the gaps have all the same values.
Interestingly the parity symmetry for  $\Omega = 0$ is unbroken in the superconducting properties since even and odd gaps are equal. Nevertheless for $\Omega \neq 0$ this degeneracy is lifted since even and odd gaps are no longer equal.
Interestingly  just a small potential barrier  is enough to do this splitting and unveil the multigap nature of the SIS nanofilm.
The interaction splitting is small ($\sim 0.5 \%$) in comparison with the shape resonance splitting ($\sim 3.0 \% $).

The even single particle states in the range  $0<a\Omega <1$, are well described by Eq.(\ref{kazero}), which is a good approximation to the exact  solution of Eq.(\ref{kcotk}) for a weak insulating barrier range.
Fig.~\ref{multiple} shows the gaps obtained by solving Eqs.(\ref{deltai}) and (\ref{Ffunc}) with matrix elements Eqs.(\ref{odd-odd}), (\ref{even-odd}), (\ref{even-even1}), and (\ref{even-even2}).  The even wavenumbers are obtained from Eq.(\ref{kainfty}), which provides an approximate solution of Eq.(\ref{kcotk}).

Fig.~\ref{delta-a-omega} shows the gaps within the same
half width window of Fig.~\ref{delta-a-zero-infty}, $0.5 \le a(nm) \le 2.5$, but for $\Omega=90 \; \mbox{nm}$. Its understanding must be complemented by Table~\ref{tab8},
which provides further information about the number of gaps and their range.
Notice that for the SIS nanofilm the number of distinct gaps is the same as the number of accessible bands.
Table~\ref{tab8} shows the critical half widths where there is a change in the number of even and odd bands within the studied range $0.5 \le a(nm) \le 2.5$.
We conclude from Fig.~\ref{delta-a-omega} that shape resonances still exist for finite $\Omega$.
Nevertheless the shape resonance acquires a more elaborate structure since the number of  even and odd single particle levels, which here is the same as the number of distinct gaps, change in similar but not equal widths.
According to Table~\ref{tab8} the tip of the curves in  Fig.~\ref{delta-a-omega}, associated to the highest possible gaps, have an equal number of even and odd single particle states below the chemical potential. At the bottom of these curves, where the lowest gap values of these curves is reached, one observes the increase by one in the number of even single particle states, according to Table~\ref{tab8}.
Next and within a very close half width value, a new transition takes place where a new odd level is added. Lastly there is a third transition corresponding to the final jump to the top of the next curve where an even level is added. Then even and odd levels become equal again.

\section{Interaction driven resonance}
In this section two Nb SIS nanofilms with fixed half widths are studied, namely  $a=$ 1.49 and  to 1.05 nm, in different ranges of $\Omega$.
The goal is to show the presence of interaction driven resonances which means that the superconducting gaps undergo abrupt changes by adjustment of $\Omega$ without changing the half width $a$.

The shape resonance, obtained by tuning of the half width, $a$, affects all discrete states but the interaction resonance can only affect the even states because only them feel the barrier.
Therefore the cause of an interaction resonance is the passage of an even single particle state through the Debye energy window that renders a new (even) band accessible for pairing.
Fig.~\ref{a_149} and Fig.~\ref{a_105} show the gaps obtained by solving Eqs.(\ref{deltai}) and (\ref{Ffunc}) with matrix elements Eqs.(\ref{odd-odd}), (\ref{even-odd}), (\ref{even-even1}), and (\ref{even-even2}). It remains to describe how the wavenumbers of Eq.(\ref{kcotk}) are obtained.
Below we study two examples of interaction driven resonance that take place in distinct ranges of barrier strength.

The first example is for $a=$ 1.49 nm  at  a fixed set of ten insulating barrier strengths, given by $a\Omega = 1,2,\ldots 10$, whose wavenumbers  are listed in Table~\ref{tab7} and correspond to exact solutions of Eq.(\ref{kcotk}).
Notice that in Table~\ref{tab7} the single particle energy level $j=7$ lies  above the Debye energy window for any $\Omega a$ in this table and for this reason does not contribute to the superconducting state. The level $j=6$ falls  below (or inside) the Debye energy window from $\Omega a =$ $1.0$ to $\Omega a =6.0$, and above from $7.0$ to $10.0$. $\Omega a =6.0$ corresponds to the barrier strength  $\Omega''$ shown in Table~\ref{tab5}.
The single particle energy level $j=5$ lies  below the Debye energy window for any $\Omega a$ in this table.
Within this range of solutions the interaction resonance can also be observed for other half widths in the proximity of $a=$ 1.49 nm.
The interaction resonance moves out of this fixed $a\Omega$ window for values of $a$ significantly different from the above one.
Fig.~\ref{a_149} shows the gaps, as obtained from Eq.(\ref{deltai}), versus the insulating barrier strength. An abrupt change of the gaps take place near to $\Omega=$ 4.03 nm$^{-1}$ which characterizes the interaction resonance.
Notice in this figure the multigap structure with odd gaps (triangles, red) above the even gaps (upside triangle, green).
Table~\ref{tab5} is complementary to Fig.~\ref{a_149} as it shows at selected values of $\Omega$, the energies of the single particle states through this transition.
By increasing the barrier strength the number of single particle states inside and below the Debye energy window changes from 6 to 5.
Specific values of $\Omega$ were selected to characterize the transition, namely, $\Omega=$ 3.36 (below),   4.03 (in)  and 4.70 (above) nm$^{-1}$.
The odd single particle energy levels are not affected by changes in the barrier strength, as shown in Table~\ref{tab5}.
Fig.~\ref{a_149} shows that the change in the gap values through the interaction resonance can be quite high ($\sim 20 \%$).

The second  example is for half width $a=$ 1.05 nm  and takes place at the range of potential barrier strength $50.0<a\Omega < 200$.
In this example the even single particle states are obtained as approximate solutions of Eq.(\ref{kcotk}) given by Eq.(\ref{kainfty}).
Fig.~\ref{a_105} shows the multigaps versus $\Omega$ with significant changes in the gap at the interaction resonance ($\sim 40 \%$).
The interaction  resonance  observed at $\Omega$ = 85.71 nm$^{-1}$ also exist in a different $\Omega$ but still within the studied range for similar half widths.
Notice that the studied range falls below $\Omega_c \sim$ 246 nm$^{-1}$ (Table~\ref{tab2})  beyond which crossing pairing between the even and odd levels must be included, which is not done here.
Table~\ref{tab6} shows through three selected values of $\Omega$, 76.19, 85.71 and 95.24 that the number of even single particle states under the chemical potential drops from 4 to 3 in this transition.
Nevertheless the number of odd levels remains constant throughout this transition.
\begin{table*}[htb]
\centering
\begin{tabular}{|c||c||c||c||c||c||c||c||c||c||c||c|}
\hline
$j$ &$j\pi$ & $\Omega a=$ 1.0 & $\Omega a=$  2.0 & $\Omega a=$ 3.0 & $\Omega a=$  4.0 &  $\Omega a=$ 5.0 &  $\Omega a=$ 6.0 &  $\Omega a=$ 7.0 &  $\Omega a=$ 8.0 &  $\Omega a=$ 9.0 & $\Omega a=$ 10.0 \\
\hline
\hline 1 &3.1416 &  2.02876 &  2.28893 &  2.45564 &  2.57043 &  2.65366	&  2.71646 &  2.76536 &  2.80443 & 2.8363	&  2.86277\\
\hline 2 &6.2832 &  4.91318	&  5.08699 &  5.23294 &  5.35403 &  5.45435	&  5.53782 &  5.60777 &  5.66687 & 5.71725      &  5.76056 \\
\hline 3 &9.4248 &  7.97867	&  8.09616 &  8.20453 &  8.30293 &  8.39135	&  8.47029 &  8.54057 &  8.60307 & 8.6587	&  8.70831 \\
\hline 4 &12.5664 & 11.0855	& 11.1727  & 11.256   & 11.3348  & 11.4086	& 11.4773  & 11.5408  & 11.5993  & 11.6532	& 11.7027 \\
\hline 5 & 15.7080 & 14.2074	& 14.2764  & 14.3434  & 14.408   & 14.4699	& 14.5288  & 14.5847  & 14.6374  & 14.6869	& 14.7335 \\
\hline 6 &18.8496 & 17.3364	& 17.3932  & 17.449   & 17.5034  & 17.5562	& 17.6072  & 17.6562  & 17.7032  & 17.7481	& 17.7908 \\
\hline 7 &21.9911 & 20.4692	& 20.5175  & 20.5652  & 20.612   & 20.6578	& 20.7024  & 20.7458  & 20.7877  & 20.8282	& 20.8672 \\
\hline
\end{tabular}
\caption{Numerical solutions of  Eq.(\ref{kcotk}) for a discrete set of barrier strength, $a\Omega=1,2, \cdots, 10$, are shown here. They are used to obtain the superconducting gaps shown in Fig.~\ref{a_149}. The solutions of Eq.(\ref{kcotk}) are displayed here as the product of the wavenumber times the half width, $ka$.
They provide the wavenumbers of the even levels while odd ones, $ka=j\pi$ are listed in the first row for comparison.
As $a\Omega$ increases even and odd solutions approach each other, as shown here.
 }
\label{tab7}
\end{table*}
\begin{table}[htb]
\centering
\begin{tabular}{|c||c||c||c|}
\hline $a=$1.49  & $\Omega=$ 3.36  & $\Omega'=$  4.03  & $\Omega''=$  4.70 \\
nm & nm$^{-1}$ & nm$^{-1}$ & nm$^{-1}$ \\
\hline
symmetry&$\epsilon_j(\Omega)/e_D$&$\epsilon_j(\Omega')/e_D$&$\epsilon_j(\Omega'')/e_D$ \\
\hline \hline
{\it even}  &5.082 \,&  5.325\, & 5.518 \,\\
\hline
{\it even} & 21.47 & 22.13 & 22.69\\
\hline
{\it even} & 50.81 & 51.77 & 52.63\\
\hline
{\it even} & 93.92& 95.06 & 96.11\\
\hline
{\it even} & 151.1& 152.3& 153.5\\
\hline \hline
{\it even} & 222.4 & 223.7 & \\
\hline \hline
{\it odd} & 7.122 \, & 7.122 \, & 7.122\\
\hline
{\it odd} & 28.49  & 28.49 & 28.49\\
\hline
{\it odd} & 64.01 & 64.01 & 64.01\\
\hline
{\it odd} & 114.0 & 114.0 & 114.0 \\
\hline
{\it odd} & 178.1 & 178.1 & 178.1\\
\hline
\end{tabular}
\caption{The single particle energy levels are shown for an SIS nanofilm with half widt $a=$ 1.49 nm and three values of $\Omega$ taken within the range $a\Omega = 1,2,\ldots, 10$.
The extremities of the Debye energy window are $\mu/e_D-1=222.9$, $\mu/e_D+1=224.9$,  according to  Eq.(\ref{debye}) and  Table~\ref{tab2}.
For $\Omega$ and $\Omega'$ there are 6 single particle energy even levels inside and below the Debye energy window
For $\Omega''$ this number drops to 5
thus  leading to an interaction driven resonance  associated to an even state.
Fig.~\ref{a_149} shows  the gaps through this transition.
Table~\ref{tab7} gives the wave numbers of these single particle energy states.
Notice that the odd levels remain unchanged throughout the transition.
}
\label{tab5}
\end{table}
\begin{table}[htb]
\centering
\begin{tabular}{|c||c||c||c|}
\hline $a=$1.05  & $\Omega=$ 76.19  & $\Omega'=$  85.71  & $\Omega''=$  95.24 \\
nm & nm$^{-1}$ & nm$^{-1}$ & nm$^{-1}$ \\
\hline
symmetry&$\epsilon_j(\Omega)/e_D$&$\epsilon_j(\Omega')/e_D$&$\epsilon_j(\Omega'')/e_D$ \\
\hline \hline
{\it even}  &13.99 \,&  14.03\, & 15.06 \,\\
\hline
{\it even} & 55.96  & 56.11  & 56.23\\
\hline
{\it even} & 125.9 & 126.3 & 126.5\\
\hline \hline
{\it even} & 223.8 & 224.4 & \\
\hline \hline
{\it odd} &  14.34 &  14.34 & 14.34\\
\hline
{\it odd} & 57.37 & 57.37 & 57.37\\
\hline
{\it odd} & 129.1 &  129.1 & 129.1\\
\hline
\end{tabular}
\caption{
Single particle energy levels are shown for $a=$ 1.05 nm and three values of $\Omega$ taken within the range $50.0 <a\Omega <200.0$. For $\Omega''$ the Debye energy window  (Table~\ref{tab2}, $\mu/e_D-1=222.9$, $\mu/e_D+1=224.9$ ) lies above the highest even energy level thus  excluding it and leading to  an interaction driven resonance.
Notice that the three odd levels remain unchanged throughout the transition.
Fig.~\ref{a_105} provides information about the gaps through this transition.
}
\label{tab6}
\end{table}

\section{Discussion}
The study of nanostructured superconductors is important for many reasons, among them the quest to predict and realize
metallic heterostructures that are multiband and multigap and present BCS-BEC crossover from momentum to position space pairing
that can be tuned by system parameters.
Theoretical and experimental evidences have shown that nanostructuring of bulk superconductors, in the forms of nanofilms,
nanostripes, and nanoclusters is able to induce multigap and multiband superconductivity and superconducting shape resonances~\cite{perali96,bianconi97,bianconi98,zhang10}. Ref.~\onlinecite{shanenko15} contains an overview of the experimental state of the art for superconducting nanofilms.
Multiband and multigap superconductors bring new paradigms to the understanding of coherent quantum phenomena, such as the enhancement of the superconducting critical temperature and the pairing energy gaps~\cite{milosevic15} through tunable parameters.
The BCS-BEC crossover is also a central issue in present studies of high-T$_c$ superconductivity. It has been proposed in Aluminum doped MgB$_2$~\cite{innocenti10} through a two-band, two-gap resonant superconductivity model.
The BCS-BEC crossover has been studied in two-band and two-gap superconductors~\cite{perali14} and also in quantum confined superconductors and superfluids~\cite{guidini16,shanenko12}.
Mostly important for connections with the results of the present work is the prediction that superconducting nanofilms can show BEC (molecule like) pairing induced by quantum confinement when the number of monolayer is reduced to a few units~\cite{chen12}.
From the other side the shape resonances in superconducting gaps and critical temperature are expected to be generally present in a quasi two-dimensional electron gas formed at oxide-oxide interface~\cite{bianconi14}, which is another system that could be used to realize the SIS structure investigated in this paper. Finally, we note that the shape resonances in superconducting nanofilms discussed at length in this paper are always accompanied by a topological change in the geometry of the Fermi surfaces, which is a Lifshitz transition.
The Lifshitz transitions seem to be a general feature of high-T$_c$ superconductors, in particular in iron-based
systems. Key predictions associated with the Lifshitz transitions are resonances and amplifications in the superconducting critical
temperature, as recently observed in electron-doped FeSe monolayer~\cite{shi16}
and a typical chemical potential (density) dependence of the superconducting isotope effect in the critical temperature~\cite{perali12}.

\section{Conclusion} \label{conclusion}
In this paper we have shown that the presence of an insulating potential barrier between two superconducting slabs in the quantum size regime brings novel and interesting effects not found in the single superconducting slab nanofilm.
This is because of the parity symmetry brought into play by the insulating barrier.
The single particle states  either vanish at the insulating barrier (odd) or not  (even). The even states are strongly affected by the insulating barrier whereas the odd ones are not, and this leads to a segregation between even and odd superconducting gaps. Consequently noticeable effects are found in the superconducting properties of the nanofilm, such as intrinsically multigapped superconductivity and interaction driven resonances, the latter being distinct from the well-known Thompson-Blatt shape resonance.
The results presented in this paper have been applied to a Nb-I-Nb nanofilm but the conclusions of multigap superconductivity and interaction driven resonances remain valid for other metals as well.

\textbf{Acknowledgments}:
We thank Antonio Bianconi, Mihaill Croitoru, Pierbiagio Pieri for useful discussions.
We also thank Natascia De Leo, Matteo Fretto and Nicola Pinto for discussions on the possible experimental realizations of SIS nanofilms.
Mauro M. Doria acknowledges CNPq support from funding (23079.014992/2015-39).
Marco Cariglia acknowledges CNPq support from funding (207007/2014-4).
Andrea Perali  acknowledges support by the University of Camerino FAR project CESEMN.
The authors thank the colleagues involved in the MultiSuper International
Network (http://www.multisuper.org) for exchange of ideas and suggestions
for this work.

\appendix
\section{Solution for the Schroedinger equation}\label{schroedinger}
Consider the one-dimensional Schroedinger equation of Eq.(\ref{schro-delta}) with the potential,
\begin{eqnarray}
U(x) = \left\{
\begin{array}{rcl}
\infty,& \mbox{if} & |x| \ge a\\
\frac{\hbar^2}{m}\Omega \, \delta \left( x \right), & \mbox{if} & |x|<a\\
\end{array}
\right.
\end{eqnarray}
One of the advantages of the delta function potential is that the energy levels are those of a free particle,
although the presence of the potential barrier,
\begin{eqnarray}
\epsilon_k = \frac{(\hbar k)^2}{2m}.
\end{eqnarray}
There are two families of solutions according to their parity properties.\\

\subsection{Even solutions: $\psi_k(-x)=\psi_k(x)$}
\begin{eqnarray}
& \psi_k(x) = \frac{1}{\sqrt{a}\sqrt{1-\frac{\sin 2 k a}{2 k a}}}\cdot \nonumber \\
&\left\{
\begin{array}{rcl}
 \sin k(x-a), & 0 < x \le a\\
-\sin k(x+a), & -a \le x <a\\
\end{array}
\right.
\end{eqnarray}
where the wavenumber is determined from Eq.(\ref{kcotk}), repeated here for clarity,
\begin{eqnarray}
ka \cdot \cot ka =- \Omega a. \nonumber
\end{eqnarray}
\subsection{Odd solutions: $\psi_k(-x)=-\psi_k(x)$}
\begin{eqnarray}
& \psi_k(x) = \frac{1}{\sqrt{a}} \cdot \nonumber \\
&\left\{
\begin{array}{rcl}
 \sin k(x-a), & 0 < x \le a\\
 \sin k(x+a), & -a \le x <a\\
\end{array}
\right.
\end{eqnarray}
where the wavenumber is given by,
\begin{eqnarray}
k a=j\pi, \quad j=1,2,3,\ldots \nonumber
\end{eqnarray}
\subsection{$\Omega a = 0$} \label{omega0}
The {\it even states} correspond to the energy levels,
\begin{eqnarray}\label{ejzeroeven}
\epsilon_j = (2j+1)^2\left ( \frac{\pi}{2a}\right)^2\frac{\hbar^2}{2m},
\end{eqnarray}
and the eigenstates in the interval $\quad |x| \leq a$ are given by
\begin{eqnarray}
\psi_j (x) = \frac{(-1)^{j+1}}{\sqrt{a}}\cos\left [ \frac{\left( j+1/2\right)\pi}{a}x\right ].
\end{eqnarray}

The {\it odd states} correspond to the energy levels,
\begin{eqnarray}\label{ejzeroodd}
\epsilon_j = j^2\left ( \frac{\pi}{a}\right)^2\frac{\hbar^2}{2m},
\end{eqnarray}
and the eigenstates in the interval $\quad |x| \leq a$ are given by
\begin{eqnarray}
\psi_j (x) = \frac{(-1)^{j}}{\sqrt{a}}\sin\left [ \frac{ j\pi}{a}x\right ].
\end{eqnarray}
\subsection{$\Omega a = \infty$} \label{omegai}
Both the {\it even states} and {\it odd states} correspond to the energy levels,
\begin{eqnarray}\label{ejinfty}
\epsilon_j = j^2\left ( \frac{\pi}{a}\right)^2\frac{\hbar^2}{2m}.
\end{eqnarray}
However the eigenstates are slightly different, while for the  {\it odd states} in the interval $\quad |x| \leq a$
correspond to
\begin{eqnarray}
\psi_j (x) = \frac{(-1)^{j}}{\sqrt{a}}\sin\left [ \frac{ j\pi}{a}x\right ],
\end{eqnarray}
for the even states is given by,
\begin{eqnarray}
& \psi_j(x) = \frac{(-1)^j}{\sqrt{a}} \cdot \nonumber \\
&\left\{
\begin{array}{rcl}
 \sin\left [ \frac{ j\pi}{a}x\right ] , & 0 < x \le a\\
 - \sin\left [ \frac{ j\pi}{a}x\right ], & -a \le x <a.\\
\end{array}
\right.
\end{eqnarray}

\section{Approximate solutions for $ka \cdot \cot ka =- \Omega a$}
We obtain useful approximated solutions of Eq.(\ref{kcotk}) valid in the limit of a weak and of a very strong potential barrier.
\subsection{$\Omega a \rightarrow 0$}
For $\Omega =0$ the solution of  Eq.(\ref{kcotk}) is $k_0a=(n+1/2)\pi$. Expressing the general solution as $ka=k_0a+\eta$ transforms Eq.(\ref{kcotk}) into $(k_0a+\eta)tan(\eta)=\Omega a$. Assuming that $\eta$ is small, $\tan (\eta) \approx \eta$ and one obtains that,
\begin{eqnarray} \label{kazero}
ka \approx \left ( n+\frac{1}{2}\right)+\sqrt{\left ( n+\frac{1}{2}\right)^2+\Omega a}.
\end{eqnarray}
\subsection{$\Omega a \rightarrow \infty$}
For $\Omega a=\infty$ the solution of  Eq.(\ref{kcotk}) is $k_0a=n\pi$.
 Expressing the general solution as $ka=k_0a+\eta$ transforms Eq.(\ref{kcotk}) into $(k_0a+\eta)/tan(\eta)=-\Omega a$.
 Similarly to the previous case, assuming that $\eta$ is small, $\tan (\eta) \approx \eta$ and one obtains that,
\begin{eqnarray}\label{kainfty}
k a \approx j \pi \frac{\Omega a}{1+\Omega a}.
\end{eqnarray}

\section{Matrix elements} \label{matrix}
Important to the present discussion is the obtainment of the {\it dimensionless} matrix elements defined in Eq.(\ref{vij}), or equivalently,
\begin{eqnarray}
V_{k\,k'} \equiv 2a\int^a_{-a}dx \, |\psi_k(x)|^2 \, |\psi_{k'}(x)|^2. \nonumber
\end{eqnarray}
The matrix element can be generally expressed as
\begin{eqnarray}\label{vkkp}
&& V_{k\,k'}= \frac{1}{\sqrt{1-\frac{\sin 2 k a}{2 k a}}}\cdot \frac{1}{\sqrt{1-\frac{\sin 2 {k'} a}{2 {k'} a}}}\cdot \nonumber \\
&& \left \{1-\left [\frac{\sin 2 k a}{2 k a}+ \frac{\sin 2 {k'} a}{2 {k'} a}\right ]+ \right . \nonumber \\
&& \left . \frac{1}{2}\left [ \frac{\sin \left ( 2( k-{k'}) a\right )}{2 (k- {k'}) a} +
\frac{\sin \left ( 2( k+{k'}) a\right )}{2 (k+ {k'}) a}
\right ] \right \}
\end{eqnarray}
From the above formula we obtain specific matrix elements.\\
\begin{enumerate}[(a)]
\item { \bf $V_{k\,k'}$,  $k$ odd, $k'$ odd }
\begin{eqnarray}\label{odd-odd}
V_{k\,k'}= 1 + \frac{1}{2} \delta_{k\,k'}.
\end{eqnarray}
The strength of the potential barrier does not affect the odd-odd matrix elements. \\
\item {\bf $V_{k\,k'}=V_{k'\,k}$,  $k$ even, $k'$ odd } \\
\begin{eqnarray}\label{even-odd}
&V_{k\,k'}=\frac{1}{1+\frac{\Omega a}{(ka)^2+(\Omega a)^2}}\cdot \nonumber \\
& \left [ 1- \frac{\Omega a}{(ka)^2+(\Omega a)^2}\cdot \frac{(k'a)^2}{(ka)^2-(k'a)^2}\right ].
\end{eqnarray}
In both limits, $\Omega \rightarrow 0$ and $\Omega \rightarrow \infty$, the formula above gives the above limit, namely, $V_{k\,k'} \rightarrow 1 + \delta_{k\,k'}/2$.
\item {\bf$V_{k\,k}$, $k$ even }\\
\begin{eqnarray}\label{even-even1}
&V_{k\,k}=\left [\frac{1}{1+\frac{\Omega a}{(ka)^2+(\Omega a)^2}} \right ]^2\cdot \nonumber \\
& \left [\frac{3}{2}\left( 1+ \frac{\Omega a}{(ka)^2+(\Omega a)^2}\right)+ \frac{(\Omega a)\cdot(ka)^2}{(ka)^2+(\Omega a)^2}\right ].
\end{eqnarray}
In both limits, $\Omega \rightarrow 0$ and $\Omega \rightarrow \infty$, the formula above gives that  $V_{k\,k'} \rightarrow 3/2$.\\
\item{\bf $V_{k\,k'}$, $k\neq k'$, $k$ even, $k'$ even }\\
\begin{eqnarray}\label{even-even2}
&V_{k\,k'}=\left [\frac{1}{1+\frac{\Omega a}{(ka)^2+(\Omega a)^2}} \right ]\cdot
\left [\frac{1}{1+\frac{\Omega a}{(k'a)^2+(\Omega a)^2}} \right ] \cdot \nonumber \\
& \left \{ 1+ (\Omega a)\frac{(ka)^2+(k'a)^2+(\Omega a)^2 }{\left[(ka)^2+(\Omega a)^2\right ]\cdot \left[(k'a)^2+(\Omega a)^2\right ]} \right \}.
\end{eqnarray}
In both limits, $\Omega \rightarrow 0$ and $\Omega \rightarrow \infty$, the formula above gives that  $V_{k\,k'} \rightarrow 1 $.
\end{enumerate}
\bibliography{reference}

\begin{thebibliography}{42}%
\makeatletter
\providecommand \@ifxundefined [1]{%
 \@ifx{#1\undefined}
}%
\providecommand \@ifnum [1]{%
 \ifnum #1\expandafter \@firstoftwo
 \else \expandafter \@secondoftwo
 \fi
}%
\providecommand \@ifx [1]{%
 \ifx #1\expandafter \@firstoftwo
 \else \expandafter \@secondoftwo
 \fi
}%
\providecommand \natexlab [1]{#1}%
\providecommand \enquote  [1]{``#1''}%
\providecommand \bibnamefont  [1]{#1}%
\providecommand \bibfnamefont [1]{#1}%
\providecommand \citenamefont [1]{#1}%
\providecommand \href@noop [0]{\@secondoftwo}%
\providecommand \href [0]{\begingroup \@sanitize@url \@href}%
\providecommand \@href[1]{\@@startlink{#1}\@@href}%
\providecommand \@@href[1]{\endgroup#1\@@endlink}%
\providecommand \@sanitize@url [0]{\catcode `\\12\catcode `\$12\catcode
  `\&12\catcode `\#12\catcode `\^12\catcode `\_12\catcode `\%12\relax}%
\providecommand \@@startlink[1]{}%
\providecommand \@@endlink[0]{}%
\providecommand \url  [0]{\begingroup\@sanitize@url \@url }%
\providecommand \@url [1]{\endgroup\@href {#1}{\urlprefix }}%
\providecommand \urlprefix  [0]{URL }%
\providecommand \Eprint [0]{\href }%
\providecommand \doibase [0]{http://dx.doi.org/}%
\providecommand \selectlanguage [0]{\@gobble}%
\providecommand \bibinfo  [0]{\@secondoftwo}%
\providecommand \bibfield  [0]{\@secondoftwo}%
\providecommand \translation [1]{[#1]}%
\providecommand \BibitemOpen [0]{}%
\providecommand \bibitemStop [0]{}%
\providecommand \bibitemNoStop [0]{.\EOS\space}%
\providecommand \EOS [0]{\spacefactor3000\relax}%
\providecommand \BibitemShut  [1]{\csname bibitem#1\endcsname}%
\let\auto@bib@innerbib\@empty
\bibitem [{\citenamefont {Liu}\ \emph {et~al.}(2001)\citenamefont {Liu},
  \citenamefont {Mazin},\ and\ \citenamefont {Kortus}}]{liu01}%
  \BibitemOpen
  \bibfield  {author} {\bibinfo {author} {\bibfnamefont {A.~Y.}\ \bibnamefont
  {Liu}}, \bibinfo {author} {\bibfnamefont {I.~I.}\ \bibnamefont {Mazin}}, \
  and\ \bibinfo {author} {\bibfnamefont {J.}~\bibnamefont {Kortus}},\ }\href
  {\doibase 10.1103/PhysRevLett.87.087005} {\bibfield  {journal} {\bibinfo
  {journal} {Phys. Rev. Lett.}\ }\textbf {\bibinfo {volume} {87}},\ \bibinfo
  {pages} {087005} (\bibinfo {year} {2001})}\BibitemShut {NoStop}%
\bibitem [{\citenamefont {Giubileo}\ \emph {et~al.}(2001)\citenamefont
  {Giubileo}, \citenamefont {Roditchev}, \citenamefont {Sacks}, \citenamefont
  {Lamy}, \citenamefont {Thanh}, \citenamefont {Klein}, \citenamefont
  {Miraglia}, \citenamefont {Fruchart}, \citenamefont {Marcus},\ and\
  \citenamefont {Monod}}]{giubileo01}%
  \BibitemOpen
  \bibfield  {author} {\bibinfo {author} {\bibfnamefont {F.}~\bibnamefont
  {Giubileo}}, \bibinfo {author} {\bibfnamefont {D.}~\bibnamefont {Roditchev}},
  \bibinfo {author} {\bibfnamefont {W.}~\bibnamefont {Sacks}}, \bibinfo
  {author} {\bibfnamefont {R.}~\bibnamefont {Lamy}}, \bibinfo {author}
  {\bibfnamefont {D.~X.}\ \bibnamefont {Thanh}}, \bibinfo {author}
  {\bibfnamefont {J.}~\bibnamefont {Klein}}, \bibinfo {author} {\bibfnamefont
  {S.}~\bibnamefont {Miraglia}}, \bibinfo {author} {\bibfnamefont
  {D.}~\bibnamefont {Fruchart}}, \bibinfo {author} {\bibfnamefont
  {J.}~\bibnamefont {Marcus}}, \ and\ \bibinfo {author} {\bibfnamefont
  {P.}~\bibnamefont {Monod}},\ }\href {\doibase 10.1103/PhysRevLett.87.177008}
  {\bibfield  {journal} {\bibinfo  {journal} {Phys. Rev. Lett.}\ }\textbf
  {\bibinfo {volume} {87}},\ \bibinfo {pages} {177008} (\bibinfo {year}
  {2001})}\BibitemShut {NoStop}%
\bibitem [{\citenamefont {Giubileo}\ \emph {et~al.}(2002)\citenamefont
  {Giubileo}, \citenamefont {Roditchev}, \citenamefont {Sacks}, \citenamefont
  {Lamy},\ and\ \citenamefont {Klein}}]{giubileo02}%
  \BibitemOpen
  \bibfield  {author} {\bibinfo {author} {\bibfnamefont {F.}~\bibnamefont
  {Giubileo}}, \bibinfo {author} {\bibfnamefont {D.}~\bibnamefont {Roditchev}},
  \bibinfo {author} {\bibfnamefont {W.}~\bibnamefont {Sacks}}, \bibinfo
  {author} {\bibfnamefont {R.}~\bibnamefont {Lamy}}, \ and\ \bibinfo {author}
  {\bibfnamefont {J.}~\bibnamefont {Klein}},\ }\href
  {http://stacks.iop.org/0295-5075/58/i=5/a=764} {\bibfield  {journal}
  {\bibinfo  {journal} {EPL (Europhysics Letters)}\ }\textbf {\bibinfo {volume}
  {58}},\ \bibinfo {pages} {764} (\bibinfo {year} {2002})}\BibitemShut
  {NoStop}%
\bibitem [{\citenamefont {Iavarone}\ \emph {et~al.}(2002)\citenamefont
  {Iavarone}, \citenamefont {Karapetrov}, \citenamefont {Koshelev},
  \citenamefont {Kwok}, \citenamefont {Crabtree}, \citenamefont {Hinks},
  \citenamefont {Kang}, \citenamefont {Choi}, \citenamefont {Kim},
  \citenamefont {Kim},\ and\ \citenamefont {Lee}}]{iavarone02}%
  \BibitemOpen
  \bibfield  {author} {\bibinfo {author} {\bibfnamefont {M.}~\bibnamefont
  {Iavarone}}, \bibinfo {author} {\bibfnamefont {G.}~\bibnamefont
  {Karapetrov}}, \bibinfo {author} {\bibfnamefont {A.~E.}\ \bibnamefont
  {Koshelev}}, \bibinfo {author} {\bibfnamefont {W.~K.}\ \bibnamefont {Kwok}},
  \bibinfo {author} {\bibfnamefont {G.~W.}\ \bibnamefont {Crabtree}}, \bibinfo
  {author} {\bibfnamefont {D.~G.}\ \bibnamefont {Hinks}}, \bibinfo {author}
  {\bibfnamefont {W.~N.}\ \bibnamefont {Kang}}, \bibinfo {author}
  {\bibfnamefont {E.-M.}\ \bibnamefont {Choi}}, \bibinfo {author}
  {\bibfnamefont {H.~J.}\ \bibnamefont {Kim}}, \bibinfo {author} {\bibfnamefont
  {H.-J.}\ \bibnamefont {Kim}}, \ and\ \bibinfo {author} {\bibfnamefont
  {S.~I.}\ \bibnamefont {Lee}},\ }\href {\doibase
  10.1103/PhysRevLett.89.187002} {\bibfield  {journal} {\bibinfo  {journal}
  {Phys. Rev. Lett.}\ }\textbf {\bibinfo {volume} {89}},\ \bibinfo {pages}
  {187002} (\bibinfo {year} {2002})}\BibitemShut {NoStop}%
\bibitem [{\citenamefont {Iavarone}\ \emph {et~al.}(2005)\citenamefont
  {Iavarone}, \citenamefont {Di~Capua}, \citenamefont {Koshelev}, \citenamefont
  {Kwok}, \citenamefont {Chiarella}, \citenamefont {Vaglio}, \citenamefont
  {Kang}, \citenamefont {Choi}, \citenamefont {Kim}, \citenamefont {Lee},
  \citenamefont {Pogrebnyakov}, \citenamefont {Redwing},\ and\ \citenamefont
  {Xi}}]{iavarone05}%
  \BibitemOpen
  \bibfield  {author} {\bibinfo {author} {\bibfnamefont {M.}~\bibnamefont
  {Iavarone}}, \bibinfo {author} {\bibfnamefont {R.}~\bibnamefont {Di~Capua}},
  \bibinfo {author} {\bibfnamefont {A.~E.}\ \bibnamefont {Koshelev}}, \bibinfo
  {author} {\bibfnamefont {W.~K.}\ \bibnamefont {Kwok}}, \bibinfo {author}
  {\bibfnamefont {F.}~\bibnamefont {Chiarella}}, \bibinfo {author}
  {\bibfnamefont {R.}~\bibnamefont {Vaglio}}, \bibinfo {author} {\bibfnamefont
  {W.~N.}\ \bibnamefont {Kang}}, \bibinfo {author} {\bibfnamefont {E.~M.}\
  \bibnamefont {Choi}}, \bibinfo {author} {\bibfnamefont {H.~J.}\ \bibnamefont
  {Kim}}, \bibinfo {author} {\bibfnamefont {S.~I.}\ \bibnamefont {Lee}},
  \bibinfo {author} {\bibfnamefont {A.~V.}\ \bibnamefont {Pogrebnyakov}},
  \bibinfo {author} {\bibfnamefont {J.~M.}\ \bibnamefont {Redwing}}, \ and\
  \bibinfo {author} {\bibfnamefont {X.~X.}\ \bibnamefont {Xi}},\ }\href
  {\doibase 10.1103/PhysRevB.71.214502} {\bibfield  {journal} {\bibinfo
  {journal} {Phys. Rev. B}\ }\textbf {\bibinfo {volume} {71}},\ \bibinfo
  {pages} {214502} (\bibinfo {year} {2005})}\BibitemShut {NoStop}%
\bibitem [{\citenamefont {Kortus}\ \emph {et~al.}(2001)\citenamefont {Kortus},
  \citenamefont {Mazin}, \citenamefont {Belashchenko}, \citenamefont
  {Antropov},\ and\ \citenamefont {Boyer}}]{kortus01}%
  \BibitemOpen
  \bibfield  {author} {\bibinfo {author} {\bibfnamefont {J.}~\bibnamefont
  {Kortus}}, \bibinfo {author} {\bibfnamefont {I.~I.}\ \bibnamefont {Mazin}},
  \bibinfo {author} {\bibfnamefont {K.~D.}\ \bibnamefont {Belashchenko}},
  \bibinfo {author} {\bibfnamefont {V.~P.}\ \bibnamefont {Antropov}}, \ and\
  \bibinfo {author} {\bibfnamefont {L.~L.}\ \bibnamefont {Boyer}},\ }\href
  {\doibase 10.1103/PhysRevLett.86.4656} {\bibfield  {journal} {\bibinfo
  {journal} {Phys. Rev. Lett.}\ }\textbf {\bibinfo {volume} {86}},\ \bibinfo
  {pages} {4656} (\bibinfo {year} {2001})}\BibitemShut {NoStop}%
\bibitem [{\citenamefont {Bianconi}\ \emph {et~al.}(2007)\citenamefont
  {Bianconi}, \citenamefont {Filippi}, \citenamefont {Fratini}, \citenamefont
  {Liarokapis}, \citenamefont {Palmisano}, \citenamefont {Saini},\ and\
  \citenamefont {Simonelli}}]{bianconi2007}%
  \BibitemOpen
  \bibfield  {author} {\bibinfo {author} {\bibfnamefont {A.}~\bibnamefont
  {Bianconi}}, \bibinfo {author} {\bibfnamefont {M.}~\bibnamefont {Filippi}},
  \bibinfo {author} {\bibfnamefont {M.}~\bibnamefont {Fratini}}, \bibinfo
  {author} {\bibfnamefont {E.}~\bibnamefont {Liarokapis}}, \bibinfo {author}
  {\bibfnamefont {V.}~\bibnamefont {Palmisano}}, \bibinfo {author}
  {\bibfnamefont {L.~N.}\ \bibnamefont {Saini}}, \ and\ \bibinfo {author}
  {\bibfnamefont {L.}~\bibnamefont {Simonelli}},\ }\enquote {\bibinfo {title}
  {Electron correlation in new materials and nanosystems},}\ \ (\bibinfo
  {publisher} {Springer Netherlands},\ \bibinfo {address} {Dordrecht},\
  \bibinfo {year} {2007})\ Chap.\ \bibinfo {chapter} {II.1 Magnesium diboride
  and the two-band scenario}, pp.\ \bibinfo {pages} {93--101}\BibitemShut
  {NoStop}%
\bibitem [{\citenamefont {Suhl}\ \emph {et~al.}(1959)\citenamefont {Suhl},
  \citenamefont {Matthias},\ and\ \citenamefont {Walker}}]{suhl1959}%
  \BibitemOpen
  \bibfield  {author} {\bibinfo {author} {\bibfnamefont {H.}~\bibnamefont
  {Suhl}}, \bibinfo {author} {\bibfnamefont {B.~T.}\ \bibnamefont {Matthias}},
  \ and\ \bibinfo {author} {\bibfnamefont {L.~R.}\ \bibnamefont {Walker}},\
  }\href {\doibase 10.1103/PhysRevLett.3.552} {\bibfield  {journal} {\bibinfo
  {journal} {Phys. Rev. Lett.}\ }\textbf {\bibinfo {volume} {3}},\ \bibinfo
  {pages} {552} (\bibinfo {year} {1959})}\BibitemShut {NoStop}%
\bibitem [{\citenamefont {Moskalenko}\ \emph {et~al.}(1991)\citenamefont
  {Moskalenko}, \citenamefont {Palistrant},\ and\ \citenamefont
  {Vakalyuk}}]{moskalenko1991}%
  \BibitemOpen
  \bibfield  {author} {\bibinfo {author} {\bibfnamefont {V.~A.}\ \bibnamefont
  {Moskalenko}}, \bibinfo {author} {\bibfnamefont {M.~E.}\ \bibnamefont
  {Palistrant}}, \ and\ \bibinfo {author} {\bibfnamefont {V.~M.}\ \bibnamefont
  {Vakalyuk}},\ }\href {http://stacks.iop.org/0038-5670/34/i=8/a=R06}
  {\bibfield  {journal} {\bibinfo  {journal} {Soviet Physics Uspekhi}\ }\textbf
  {\bibinfo {volume} {34}},\ \bibinfo {pages} {717} (\bibinfo {year}
  {1991})}\BibitemShut {NoStop}%
\bibitem [{\citenamefont {Perucchi}\ \emph {et~al.}(2012)\citenamefont
  {Perucchi}, \citenamefont {Baldassarre}, \citenamefont {Lupi}, \citenamefont
  {Joseph},\ and\ \citenamefont {Dore}}]{perucchi12}%
  \BibitemOpen
  \bibfield  {author} {\bibinfo {author} {\bibfnamefont {A.}~\bibnamefont
  {Perucchi}}, \bibinfo {author} {\bibfnamefont {L.}~\bibnamefont
  {Baldassarre}}, \bibinfo {author} {\bibfnamefont {S.}~\bibnamefont {Lupi}},
  \bibinfo {author} {\bibfnamefont {B.}~\bibnamefont {Joseph}}, \ and\ \bibinfo
  {author} {\bibfnamefont {P.}~\bibnamefont {Dore}},\ }in\ \href {\doibase
  10.1109/IRMMW-THz.2012.6379500} {\emph {\bibinfo {booktitle} {Infrared,
  Millimeter, and Terahertz Waves (IRMMW-THz), 2012 37th International
  Conference on}}}\ (\bibinfo {year} {2012})\ pp.\ \bibinfo {pages}
  {1--3}\BibitemShut {NoStop}%
\bibitem [{\citenamefont {Abdel-Hafiez}\ \emph {et~al.}(2015)\citenamefont
  {Abdel-Hafiez}, \citenamefont {Zhang}, \citenamefont {Cao}, \citenamefont
  {Duan}, \citenamefont {Karapetrov}, \citenamefont {Pudalov}, \citenamefont
  {Vlasenko}, \citenamefont {Sadakov}, \citenamefont {Knyazev}, \citenamefont
  {Romanova}, \citenamefont {Chareev}, \citenamefont {Volkova}, \citenamefont
  {Vasiliev},\ and\ \citenamefont {Chen}}]{hafiez15}%
  \BibitemOpen
  \bibfield  {author} {\bibinfo {author} {\bibfnamefont {M.}~\bibnamefont
  {Abdel-Hafiez}}, \bibinfo {author} {\bibfnamefont {Y.-Y.}\ \bibnamefont
  {Zhang}}, \bibinfo {author} {\bibfnamefont {Z.-Y.}\ \bibnamefont {Cao}},
  \bibinfo {author} {\bibfnamefont {C.-G.}\ \bibnamefont {Duan}}, \bibinfo
  {author} {\bibfnamefont {G.}~\bibnamefont {Karapetrov}}, \bibinfo {author}
  {\bibfnamefont {V.~M.}\ \bibnamefont {Pudalov}}, \bibinfo {author}
  {\bibfnamefont {V.~A.}\ \bibnamefont {Vlasenko}}, \bibinfo {author}
  {\bibfnamefont {A.~V.}\ \bibnamefont {Sadakov}}, \bibinfo {author}
  {\bibfnamefont {D.~A.}\ \bibnamefont {Knyazev}}, \bibinfo {author}
  {\bibfnamefont {T.~A.}\ \bibnamefont {Romanova}}, \bibinfo {author}
  {\bibfnamefont {D.~A.}\ \bibnamefont {Chareev}}, \bibinfo {author}
  {\bibfnamefont {O.~S.}\ \bibnamefont {Volkova}}, \bibinfo {author}
  {\bibfnamefont {A.~N.}\ \bibnamefont {Vasiliev}}, \ and\ \bibinfo {author}
  {\bibfnamefont {X.-J.}\ \bibnamefont {Chen}},\ }\href {\doibase
  10.1103/PhysRevB.91.165109} {\bibfield  {journal} {\bibinfo  {journal} {Phys.
  Rev. B}\ }\textbf {\bibinfo {volume} {91}},\ \bibinfo {pages} {165109}
  (\bibinfo {year} {2015})}\BibitemShut {NoStop}%
\bibitem [{\citenamefont {Floris}\ \emph {et~al.}(2007)\citenamefont {Floris},
  \citenamefont {Sanna}, \citenamefont {Massidda},\ and\ \citenamefont
  {Gross}}]{floris07}%
  \BibitemOpen
  \bibfield  {author} {\bibinfo {author} {\bibfnamefont {A.}~\bibnamefont
  {Floris}}, \bibinfo {author} {\bibfnamefont {A.}~\bibnamefont {Sanna}},
  \bibinfo {author} {\bibfnamefont {S.}~\bibnamefont {Massidda}}, \ and\
  \bibinfo {author} {\bibfnamefont {E.~K.~U.}\ \bibnamefont {Gross}},\ }\href
  {\doibase 10.1103/PhysRevB.75.054508} {\bibfield  {journal} {\bibinfo
  {journal} {Phys. Rev. B}\ }\textbf {\bibinfo {volume} {75}},\ \bibinfo
  {pages} {054508} (\bibinfo {year} {2007})}\BibitemShut {NoStop}%
\bibitem [{\citenamefont {Ruby}\ \emph {et~al.}(2015)\citenamefont {Ruby},
  \citenamefont {Heinrich}, \citenamefont {Pascual},\ and\ \citenamefont
  {Franke}}]{ruby15}%
  \BibitemOpen
  \bibfield  {author} {\bibinfo {author} {\bibfnamefont {M.}~\bibnamefont
  {Ruby}}, \bibinfo {author} {\bibfnamefont {B.~W.}\ \bibnamefont {Heinrich}},
  \bibinfo {author} {\bibfnamefont {J.~I.}\ \bibnamefont {Pascual}}, \ and\
  \bibinfo {author} {\bibfnamefont {K.~J.}\ \bibnamefont {Franke}},\ }\href
  {\doibase 10.1103/PhysRevLett.114.157001} {\bibfield  {journal} {\bibinfo
  {journal} {Phys. Rev. Lett.}\ }\textbf {\bibinfo {volume} {114}},\ \bibinfo
  {pages} {157001} (\bibinfo {year} {2015})}\BibitemShut {NoStop}%
\bibitem [{\citenamefont {Guo}\ \emph {et~al.}(2004)\citenamefont {Guo},
  \citenamefont {Zhang}, \citenamefont {Bao}, \citenamefont {Han},
  \citenamefont {Tang}, \citenamefont {Zhang}, \citenamefont {Zhu},
  \citenamefont {Wang}, \citenamefont {Niu}, \citenamefont {Qiu}, \citenamefont
  {Jia}, \citenamefont {Zhao},\ and\ \citenamefont {Xue}}]{guo04}%
  \BibitemOpen
  \bibfield  {author} {\bibinfo {author} {\bibfnamefont {Y.}~\bibnamefont
  {Guo}}, \bibinfo {author} {\bibfnamefont {Y.-F.}\ \bibnamefont {Zhang}},
  \bibinfo {author} {\bibfnamefont {X.-Y.}\ \bibnamefont {Bao}}, \bibinfo
  {author} {\bibfnamefont {T.-Z.}\ \bibnamefont {Han}}, \bibinfo {author}
  {\bibfnamefont {Z.}~\bibnamefont {Tang}}, \bibinfo {author} {\bibfnamefont
  {L.-X.}\ \bibnamefont {Zhang}}, \bibinfo {author} {\bibfnamefont {W.-G.}\
  \bibnamefont {Zhu}}, \bibinfo {author} {\bibfnamefont {E.~G.}\ \bibnamefont
  {Wang}}, \bibinfo {author} {\bibfnamefont {Q.}~\bibnamefont {Niu}}, \bibinfo
  {author} {\bibfnamefont {Z.~Q.}\ \bibnamefont {Qiu}}, \bibinfo {author}
  {\bibfnamefont {J.-F.}\ \bibnamefont {Jia}}, \bibinfo {author} {\bibfnamefont
  {Z.-X.}\ \bibnamefont {Zhao}}, \ and\ \bibinfo {author} {\bibfnamefont
  {Q.-K.}\ \bibnamefont {Xue}},\ }\href {\doibase 10.1126/science.1105130}
  {\bibfield  {journal} {\bibinfo  {journal} {Science}\ }\textbf {\bibinfo
  {volume} {306}},\ \bibinfo {pages} {1915} (\bibinfo {year}
  {2004})}\BibitemShut {NoStop}%
\bibitem [{\citenamefont {Bao}\ \emph {et~al.}(2005)\citenamefont {Bao},
  \citenamefont {Zhang}, \citenamefont {Wang}, \citenamefont {Jia},
  \citenamefont {Xue}, \citenamefont {Xie},\ and\ \citenamefont
  {Zhao}}]{bao05}%
  \BibitemOpen
  \bibfield  {author} {\bibinfo {author} {\bibfnamefont {X.-Y.}\ \bibnamefont
  {Bao}}, \bibinfo {author} {\bibfnamefont {Y.-F.}\ \bibnamefont {Zhang}},
  \bibinfo {author} {\bibfnamefont {Y.}~\bibnamefont {Wang}}, \bibinfo {author}
  {\bibfnamefont {J.-F.}\ \bibnamefont {Jia}}, \bibinfo {author} {\bibfnamefont
  {Q.-K.}\ \bibnamefont {Xue}}, \bibinfo {author} {\bibfnamefont {X.~C.}\
  \bibnamefont {Xie}}, \ and\ \bibinfo {author} {\bibfnamefont {Z.-X.}\
  \bibnamefont {Zhao}},\ }\href {\doibase 10.1103/PhysRevLett.95.247005}
  {\bibfield  {journal} {\bibinfo  {journal} {Phys. Rev. Lett.}\ }\textbf
  {\bibinfo {volume} {95}},\ \bibinfo {pages} {247005} (\bibinfo {year}
  {2005})}\BibitemShut {NoStop}%
\bibitem [{\citenamefont {Zhang}\ \emph {et~al.}(2010)\citenamefont {Zhang},
  \citenamefont {Cheng}, \citenamefont {Li}, \citenamefont {Sun}, \citenamefont
  {Wang}, \citenamefont {Zhu}, \citenamefont {He}, \citenamefont {Wang},
  \citenamefont {Ma}, \citenamefont {Chen}, \citenamefont {Wang}, \citenamefont
  {Liu}, \citenamefont {Lin}, \citenamefont {Jia},\ and\ \citenamefont
  {Xue}}]{zhang10}%
  \BibitemOpen
  \bibfield  {author} {\bibinfo {author} {\bibfnamefont {T.}~\bibnamefont
  {Zhang}}, \bibinfo {author} {\bibfnamefont {P.}~\bibnamefont {Cheng}},
  \bibinfo {author} {\bibfnamefont {W.-J.}\ \bibnamefont {Li}}, \bibinfo
  {author} {\bibfnamefont {Y.-J.}\ \bibnamefont {Sun}}, \bibinfo {author}
  {\bibfnamefont {G.}~\bibnamefont {Wang}}, \bibinfo {author} {\bibfnamefont
  {X.-G.}\ \bibnamefont {Zhu}}, \bibinfo {author} {\bibfnamefont
  {K.}~\bibnamefont {He}}, \bibinfo {author} {\bibfnamefont {L.}~\bibnamefont
  {Wang}}, \bibinfo {author} {\bibfnamefont {X.}~\bibnamefont {Ma}}, \bibinfo
  {author} {\bibfnamefont {X.}~\bibnamefont {Chen}}, \bibinfo {author}
  {\bibfnamefont {Y.}~\bibnamefont {Wang}}, \bibinfo {author} {\bibfnamefont
  {Y.}~\bibnamefont {Liu}}, \bibinfo {author} {\bibfnamefont {H.-Q.}\
  \bibnamefont {Lin}}, \bibinfo {author} {\bibfnamefont {J.-F.}\ \bibnamefont
  {Jia}}, \ and\ \bibinfo {author} {\bibfnamefont {Q.-K.}\ \bibnamefont
  {Xue}},\ }\href {\doibase 10.1103/PhysRevLett.95.247005} {\bibfield
  {journal} {\bibinfo  {journal} {Nat Phys}\ }\textbf {\bibinfo {volume} {6}},\
  \bibinfo {pages} {104} (\bibinfo {year} {2010})}\BibitemShut {NoStop}%
\bibitem [{\citenamefont {Thompson}\ and\ \citenamefont
  {Blatt}(1963)}]{thompson63}%
  \BibitemOpen
  \bibfield  {author} {\bibinfo {author} {\bibfnamefont {C.}~\bibnamefont
  {Thompson}}\ and\ \bibinfo {author} {\bibfnamefont {J.}~\bibnamefont
  {Blatt}},\ }\href {\doibase http://dx.doi.org/10.1016/S0375-9601(63)80003-1}
  {\bibfield  {journal} {\bibinfo  {journal} {Physics Letters}\ }\textbf
  {\bibinfo {volume} {5}},\ \bibinfo {pages} {6 } (\bibinfo {year}
  {1963})}\BibitemShut {NoStop}%
\bibitem [{\citenamefont {Perali}\ \emph {et~al.}(1996)\citenamefont {Perali},
  \citenamefont {Bianconi}, \citenamefont {Lanzara},\ and\ \citenamefont
  {Saini}}]{perali96}%
  \BibitemOpen
  \bibfield  {author} {\bibinfo {author} {\bibfnamefont {A.}~\bibnamefont
  {Perali}}, \bibinfo {author} {\bibfnamefont {A.}~\bibnamefont {Bianconi}},
  \bibinfo {author} {\bibfnamefont {A.}~\bibnamefont {Lanzara}}, \ and\
  \bibinfo {author} {\bibfnamefont {N.}~\bibnamefont {Saini}},\ }\href
  {\doibase http://dx.doi.org/10.1016/0038-1098(96)00373-0} {\bibfield
  {journal} {\bibinfo  {journal} {Solid State Communications}\ }\textbf
  {\bibinfo {volume} {100}},\ \bibinfo {pages} {181 } (\bibinfo {year}
  {1996})}\BibitemShut {NoStop}%
\bibitem [{\citenamefont {Shanenko}\ and\ \citenamefont
  {Croitoru}(2006)}]{shanenko06}%
  \BibitemOpen
  \bibfield  {author} {\bibinfo {author} {\bibfnamefont {A.~A.}\ \bibnamefont
  {Shanenko}}\ and\ \bibinfo {author} {\bibfnamefont {M.~D.}\ \bibnamefont
  {Croitoru}},\ }\href {\doibase 10.1103/PhysRevB.73.012510} {\bibfield
  {journal} {\bibinfo  {journal} {Phys. Rev. B}\ }\textbf {\bibinfo {volume}
  {73}},\ \bibinfo {pages} {012510} (\bibinfo {year} {2006})}\BibitemShut
  {NoStop}%
\bibitem [{\citenamefont {Josephson}(1962)}]{josephson1962}%
  \BibitemOpen
  \bibfield  {author} {\bibinfo {author} {\bibfnamefont {B.}~\bibnamefont
  {Josephson}},\ }\href {\doibase
  http://dx.doi.org/10.1016/0031-9163(62)91369-0} {\bibfield  {journal}
  {\bibinfo  {journal} {Physics Letters}\ }\textbf {\bibinfo {volume} {1}},\
  \bibinfo {pages} {251 } (\bibinfo {year} {1962})}\BibitemShut {NoStop}%
\bibitem [{\citenamefont {Golubov}\ \emph {et~al.}(2004)\citenamefont
  {Golubov}, \citenamefont {Kupriyanov},\ and\ \citenamefont
  {Il'ichev}}]{golubov76}%
  \BibitemOpen
  \bibfield  {author} {\bibinfo {author} {\bibfnamefont {A.~A.}\ \bibnamefont
  {Golubov}}, \bibinfo {author} {\bibfnamefont {M.~Y.}\ \bibnamefont
  {Kupriyanov}}, \ and\ \bibinfo {author} {\bibfnamefont {E.}~\bibnamefont
  {Il'ichev}},\ }\href {\doibase 10.1103/RevModPhys.76.411} {\bibfield
  {journal} {\bibinfo  {journal} {Rev. Mod. Phys.}\ }\textbf {\bibinfo {volume}
  {76}},\ \bibinfo {pages} {411} (\bibinfo {year} {2004})}\BibitemShut
  {NoStop}%
\bibitem [{\citenamefont {Kim}\ \emph {et~al.}(2004)\citenamefont {Kim},
  \citenamefont {Chua}, \citenamefont {Fiete}, \citenamefont {Nam},
  \citenamefont {MacDonald},\ and\ \citenamefont {Shih}}]{kim04}%
  \BibitemOpen
  \bibfield  {author} {\bibinfo {author} {\bibfnamefont {J.}~\bibnamefont
  {Kim}}, \bibinfo {author} {\bibfnamefont {V.}~\bibnamefont {Chua}}, \bibinfo
  {author} {\bibfnamefont {G.~A.}\ \bibnamefont {Fiete}}, \bibinfo {author}
  {\bibfnamefont {H.}~\bibnamefont {Nam}}, \bibinfo {author} {\bibfnamefont
  {A.~H.}\ \bibnamefont {MacDonald}}, \ and\ \bibinfo {author} {\bibfnamefont
  {C.-K.}\ \bibnamefont {Shih}},\ }\href {\doibase 10.1038/nphys2287}
  {\bibfield  {journal} {\bibinfo  {journal} {Nat Phys}\ }\textbf {\bibinfo
  {volume} {8}},\ \bibinfo {pages} {464} (\bibinfo {year} {2004})}\BibitemShut
  {NoStop}%
\bibitem [{\citenamefont {Serrier-Garcia}\ \emph {et~al.}(2013)\citenamefont
  {Serrier-Garcia}, \citenamefont {Cuevas}, \citenamefont {Cren}, \citenamefont
  {Brun}, \citenamefont {Cherkez}, \citenamefont {Debontridder}, \citenamefont
  {Fokin}, \citenamefont {Bergeret},\ and\ \citenamefont
  {Roditchev}}]{serrier13}%
  \BibitemOpen
  \bibfield  {author} {\bibinfo {author} {\bibfnamefont {L.}~\bibnamefont
  {Serrier-Garcia}}, \bibinfo {author} {\bibfnamefont {J.~C.}\ \bibnamefont
  {Cuevas}}, \bibinfo {author} {\bibfnamefont {T.}~\bibnamefont {Cren}},
  \bibinfo {author} {\bibfnamefont {C.}~\bibnamefont {Brun}}, \bibinfo {author}
  {\bibfnamefont {V.}~\bibnamefont {Cherkez}}, \bibinfo {author} {\bibfnamefont
  {F.}~\bibnamefont {Debontridder}}, \bibinfo {author} {\bibfnamefont
  {D.}~\bibnamefont {Fokin}}, \bibinfo {author} {\bibfnamefont {F.~S.}\
  \bibnamefont {Bergeret}}, \ and\ \bibinfo {author} {\bibfnamefont
  {D.}~\bibnamefont {Roditchev}},\ }\href {\doibase
  10.1103/PhysRevLett.110.157003} {\bibfield  {journal} {\bibinfo  {journal}
  {Phys. Rev. Lett.}\ }\textbf {\bibinfo {volume} {110}},\ \bibinfo {pages}
  {157003} (\bibinfo {year} {2013})}\BibitemShut {NoStop}%
\bibitem [{\citenamefont {Valentinis}\ \emph
  {et~al.}(2016{\natexlab{a}})\citenamefont {Valentinis}, \citenamefont
  {van~der Marel},\ and\ \citenamefont {C.}}]{valentinis16a}%
  \BibitemOpen
  \bibfield  {author} {\bibinfo {author} {\bibfnamefont {D.}~\bibnamefont
  {Valentinis}}, \bibinfo {author} {\bibfnamefont {D.}~\bibnamefont {van~der
  Marel}}, \ and\ \bibinfo {author} {\bibfnamefont {B.}~\bibnamefont {C.}},\
  }\href {http://arxiv.org/abs/1601.04927v1} {\bibfield  {journal} {\bibinfo
  {journal} {arXiv:1601.04927v1}\ } (\bibinfo {year}
  {2016}{\natexlab{a}})}\BibitemShut {NoStop}%
\bibitem [{\citenamefont {Valentinis}\ \emph
  {et~al.}(2016{\natexlab{b}})\citenamefont {Valentinis}, \citenamefont
  {van~der Marel},\ and\ \citenamefont {C.}}]{valentinis16b}%
  \BibitemOpen
  \bibfield  {author} {\bibinfo {author} {\bibfnamefont {D.}~\bibnamefont
  {Valentinis}}, \bibinfo {author} {\bibfnamefont {D.}~\bibnamefont {van~der
  Marel}}, \ and\ \bibinfo {author} {\bibfnamefont {B.}~\bibnamefont {C.}},\
  }\href {http://arxiv.org/abs/1601.04521} {\bibfield  {journal} {\bibinfo
  {journal} {arXiv:1601.04521}\ } (\bibinfo {year}
  {2016}{\natexlab{b}})}\BibitemShut {NoStop}%
\bibitem [{\citenamefont {Russo}\ \emph {et~al.}(2014)\citenamefont {Russo},
  \citenamefont {Granata}, \citenamefont {Vettoliere}, \citenamefont
  {Esposito}, \citenamefont {Fretto}, \citenamefont {Leo}, \citenamefont
  {Enrico},\ and\ \citenamefont {Lacquaniti}}]{russo14}%
  \BibitemOpen
  \bibfield  {author} {\bibinfo {author} {\bibfnamefont {R.}~\bibnamefont
  {Russo}}, \bibinfo {author} {\bibfnamefont {C.}~\bibnamefont {Granata}},
  \bibinfo {author} {\bibfnamefont {A.}~\bibnamefont {Vettoliere}}, \bibinfo
  {author} {\bibfnamefont {E.}~\bibnamefont {Esposito}}, \bibinfo {author}
  {\bibfnamefont {M.}~\bibnamefont {Fretto}}, \bibinfo {author} {\bibfnamefont
  {N.~D.}\ \bibnamefont {Leo}}, \bibinfo {author} {\bibfnamefont
  {E.}~\bibnamefont {Enrico}}, \ and\ \bibinfo {author} {\bibfnamefont
  {V.}~\bibnamefont {Lacquaniti}},\ }\href
  {http://stacks.iop.org/0953-2048/27/i=4/a=044028} {\bibfield  {journal}
  {\bibinfo  {journal} {Superconductor Science and Technology}\ }\textbf
  {\bibinfo {volume} {27}},\ \bibinfo {pages} {044028} (\bibinfo {year}
  {2014})}\BibitemShut {NoStop}%
\bibitem [{\citenamefont {Fl\"ugge}(2013)}]{flugge47}%
  \BibitemOpen
  \bibfield  {author} {\bibinfo {author} {\bibfnamefont {S.}~\bibnamefont
  {Fl\"ugge}},\ }\href@noop {} {\emph {\bibinfo {title} {Practical Quantum
  Mechanics}}},\ classics in mathematics\ (\bibinfo  {publisher}
  {Springer-Verlag},\ \bibinfo {year} {2013})\BibitemShut {NoStop}%
\bibitem [{\citenamefont {Kittel}(1976)}]{kittel76}%
  \BibitemOpen
  \bibfield  {author} {\bibinfo {author} {\bibfnamefont {C.}~\bibnamefont
  {Kittel}},\ }\href@noop {} {\emph {\bibinfo {title} {Introduction to Solid
  State Physics}}}\ (\bibinfo  {publisher} {John Wiley \& Sons},\ \bibinfo
  {year} {1976})\BibitemShut {NoStop}%
\bibitem [{\citenamefont {Ashcroft}\ and\ \citenamefont
  {Mermin}(1976)}]{ashcroft76}%
  \BibitemOpen
  \bibfield  {author} {\bibinfo {author} {\bibfnamefont {N.}~\bibnamefont
  {Ashcroft}}\ and\ \bibinfo {author} {\bibfnamefont {N.}~\bibnamefont
  {Mermin}},\ }\href@noop {} {\emph {\bibinfo {title} {Solid State Physics}}},\
  Science: Physics\ (\bibinfo  {publisher} {Saunders College},\ \bibinfo {year}
  {1976})\BibitemShut {NoStop}%
\bibitem [{\citenamefont {Rohlf}(1994)}]{rohlf94}%
  \BibitemOpen
  \bibfield  {author} {\bibinfo {author} {\bibfnamefont {J.~W.}\ \bibnamefont
  {Rohlf}},\ }\href@noop {} {\emph {\bibinfo {title} {Modern Physics from a to
  z}}}\ (\bibinfo  {publisher} {Wiley},\ \bibinfo {year} {1994})\BibitemShut
  {NoStop}%
\bibitem [{\citenamefont {Bianconi}\ \emph {et~al.}(1997)\citenamefont
  {Bianconi}, \citenamefont {Valletta}, \citenamefont {Perali},\ and\
  \citenamefont {Saini}}]{bianconi97}%
  \BibitemOpen
  \bibfield  {author} {\bibinfo {author} {\bibfnamefont {A.}~\bibnamefont
  {Bianconi}}, \bibinfo {author} {\bibfnamefont {A.}~\bibnamefont {Valletta}},
  \bibinfo {author} {\bibfnamefont {A.}~\bibnamefont {Perali}}, \ and\ \bibinfo
  {author} {\bibfnamefont {N.}~\bibnamefont {Saini}},\ }\href {\doibase
  http://dx.doi.org/10.1016/S0038-1098(97)00011-2} {\bibfield  {journal}
  {\bibinfo  {journal} {Solid State Communications}\ }\textbf {\bibinfo
  {volume} {102}},\ \bibinfo {pages} {369 } (\bibinfo {year}
  {1997})}\BibitemShut {NoStop}%
\bibitem [{\citenamefont {Bianconi}\ \emph {et~al.}(1998)\citenamefont
  {Bianconi}, \citenamefont {Valletta}, \citenamefont {Perali},\ and\
  \citenamefont {Saini}}]{bianconi98}%
  \BibitemOpen
  \bibfield  {author} {\bibinfo {author} {\bibfnamefont {A.}~\bibnamefont
  {Bianconi}}, \bibinfo {author} {\bibfnamefont {A.}~\bibnamefont {Valletta}},
  \bibinfo {author} {\bibfnamefont {A.}~\bibnamefont {Perali}}, \ and\ \bibinfo
  {author} {\bibfnamefont {N.~L.}\ \bibnamefont {Saini}},\ }\href {\doibase
  http://dx.doi.org/10.1016/S0921-4534(97)01825-X} {\bibfield  {journal}
  {\bibinfo  {journal} {Physica C: Superconductivity}\ }\textbf {\bibinfo
  {volume} {296}},\ \bibinfo {pages} {269 } (\bibinfo {year}
  {1998})}\BibitemShut {NoStop}%
\bibitem [{\citenamefont {Shanenko}\ \emph {et~al.}(2015)\citenamefont
  {Shanenko}, \citenamefont {Aguiar}, \citenamefont {Vagov}, \citenamefont
  {Croitoru},\ and\ \citenamefont {Milošević}}]{shanenko15}%
  \BibitemOpen
  \bibfield  {author} {\bibinfo {author} {\bibfnamefont {A.~A.}\ \bibnamefont
  {Shanenko}}, \bibinfo {author} {\bibfnamefont {J.~A.}\ \bibnamefont
  {Aguiar}}, \bibinfo {author} {\bibfnamefont {A.}~\bibnamefont {Vagov}},
  \bibinfo {author} {\bibfnamefont {M.~D.}\ \bibnamefont {Croitoru}}, \ and\
  \bibinfo {author} {\bibfnamefont {M.~V.}\ \bibnamefont {Milošević}},\
  }\href {http://stacks.iop.org/0953-2048/28/i=5/a=054001} {\bibfield
  {journal} {\bibinfo  {journal} {Superconductor Science and Technology}\
  }\textbf {\bibinfo {volume} {28}},\ \bibinfo {pages} {054001} (\bibinfo
  {year} {2015})}\BibitemShut {NoStop}%
\bibitem [{\citenamefont {Milošević}\ and\ \citenamefont
  {Perali}(2015)}]{milosevic15}%
  \BibitemOpen
  \bibfield  {author} {\bibinfo {author} {\bibfnamefont {M.~V.}\ \bibnamefont
  {Milošević}}\ and\ \bibinfo {author} {\bibfnamefont {A.}~\bibnamefont
  {Perali}},\ }\href {http://stacks.iop.org/0953-2048/28/i=6/a=060201}
  {\bibfield  {journal} {\bibinfo  {journal} {Superconductor Science and
  Technology}\ }\textbf {\bibinfo {volume} {28}},\ \bibinfo {pages} {060201}
  (\bibinfo {year} {2015})}\BibitemShut {NoStop}%
\bibitem [{\citenamefont {Innocenti}\ \emph {et~al.}(2010)\citenamefont
  {Innocenti}, \citenamefont {Poccia}, \citenamefont {Ricci}, \citenamefont
  {Valletta}, \citenamefont {Caprara}, \citenamefont {Perali},\ and\
  \citenamefont {Bianconi}}]{innocenti10}%
  \BibitemOpen
  \bibfield  {author} {\bibinfo {author} {\bibfnamefont {D.}~\bibnamefont
  {Innocenti}}, \bibinfo {author} {\bibfnamefont {N.}~\bibnamefont {Poccia}},
  \bibinfo {author} {\bibfnamefont {A.}~\bibnamefont {Ricci}}, \bibinfo
  {author} {\bibfnamefont {A.}~\bibnamefont {Valletta}}, \bibinfo {author}
  {\bibfnamefont {S.}~\bibnamefont {Caprara}}, \bibinfo {author} {\bibfnamefont
  {A.}~\bibnamefont {Perali}}, \ and\ \bibinfo {author} {\bibfnamefont
  {A.}~\bibnamefont {Bianconi}},\ }\href {\doibase 10.1103/PhysRevB.82.184528}
  {\bibfield  {journal} {\bibinfo  {journal} {Phys. Rev. B}\ }\textbf {\bibinfo
  {volume} {82}},\ \bibinfo {pages} {184528} (\bibinfo {year}
  {2010})}\BibitemShut {NoStop}%
\bibitem [{\citenamefont {Guidini}\ and\ \citenamefont
  {Perali}(2014)}]{perali14}%
  \BibitemOpen
  \bibfield  {author} {\bibinfo {author} {\bibfnamefont {A.}~\bibnamefont
  {Guidini}}\ and\ \bibinfo {author} {\bibfnamefont {A.}~\bibnamefont
  {Perali}},\ }\href {http://stacks.iop.org/0953-2048/27/i=12/a=124002}
  {\bibfield  {journal} {\bibinfo  {journal} {Superconductor Science and
  Technology}\ }\textbf {\bibinfo {volume} {27}},\ \bibinfo {pages} {124002}
  (\bibinfo {year} {2014})}\BibitemShut {NoStop}%
\bibitem [{\citenamefont {Guidini}\ \emph {et~al.}(2016)\citenamefont
  {Guidini}, \citenamefont {Flammia}, \citenamefont {Milo{\v{s}}evi{\'{c}}},\
  and\ \citenamefont {Perali}}]{guidini16}%
  \BibitemOpen
  \bibfield  {author} {\bibinfo {author} {\bibfnamefont {A.}~\bibnamefont
  {Guidini}}, \bibinfo {author} {\bibfnamefont {L.}~\bibnamefont {Flammia}},
  \bibinfo {author} {\bibfnamefont {M.~V.}\ \bibnamefont
  {Milo{\v{s}}evi{\'{c}}}}, \ and\ \bibinfo {author} {\bibfnamefont
  {A.}~\bibnamefont {Perali}},\ }\href {\doibase 10.1007/s10948-015-3308-y}
  {\bibfield  {journal} {\bibinfo  {journal} {Journal of Superconductivity and
  Novel Magnetism}\ }\textbf {\bibinfo {volume} {29}},\ \bibinfo {pages} {711}
  (\bibinfo {year} {2016})}\BibitemShut {NoStop}%
\bibitem [{\citenamefont {Shanenko}\ \emph {et~al.}(2012)\citenamefont
  {Shanenko}, \citenamefont {Croitoru}, \citenamefont {Vagov}, \citenamefont
  {Axt}, \citenamefont {Perali},\ and\ \citenamefont {Peeters}}]{shanenko12}%
  \BibitemOpen
  \bibfield  {author} {\bibinfo {author} {\bibfnamefont {A.~A.}\ \bibnamefont
  {Shanenko}}, \bibinfo {author} {\bibfnamefont {M.~D.}\ \bibnamefont
  {Croitoru}}, \bibinfo {author} {\bibfnamefont {A.~V.}\ \bibnamefont {Vagov}},
  \bibinfo {author} {\bibfnamefont {V.~M.}\ \bibnamefont {Axt}}, \bibinfo
  {author} {\bibfnamefont {A.}~\bibnamefont {Perali}}, \ and\ \bibinfo {author}
  {\bibfnamefont {F.~M.}\ \bibnamefont {Peeters}},\ }\href {\doibase
  10.1103/PhysRevA.86.033612} {\bibfield  {journal} {\bibinfo  {journal} {Phys.
  Rev. A}\ }\textbf {\bibinfo {volume} {86}},\ \bibinfo {pages} {033612}
  (\bibinfo {year} {2012})}\BibitemShut {NoStop}%
\bibitem [{\citenamefont {Chen}\ \emph {et~al.}(2012)\citenamefont {Chen},
  \citenamefont {Shanenko}, \citenamefont {Perali},\ and\ \citenamefont
  {Peeters}}]{chen12}%
  \BibitemOpen
  \bibfield  {author} {\bibinfo {author} {\bibfnamefont {Y.}~\bibnamefont
  {Chen}}, \bibinfo {author} {\bibfnamefont {A.~A.}\ \bibnamefont {Shanenko}},
  \bibinfo {author} {\bibfnamefont {A.}~\bibnamefont {Perali}}, \ and\ \bibinfo
  {author} {\bibfnamefont {F.~M.}\ \bibnamefont {Peeters}},\ }\href
  {http://stacks.iop.org/0953-8984/24/i=18/a=185701} {\bibfield  {journal}
  {\bibinfo  {journal} {Journal of Physics: Condensed Matter}\ }\textbf
  {\bibinfo {volume} {24}},\ \bibinfo {pages} {185701} (\bibinfo {year}
  {2012})}\BibitemShut {NoStop}%
\bibitem [{\citenamefont {Bianconi}\ \emph {et~al.}(2014)\citenamefont
  {Bianconi}, \citenamefont {Innocenti}, \citenamefont {Valletta},\ and\
  \citenamefont {Perali}}]{bianconi14}%
  \BibitemOpen
  \bibfield  {author} {\bibinfo {author} {\bibfnamefont {A.}~\bibnamefont
  {Bianconi}}, \bibinfo {author} {\bibfnamefont {D.}~\bibnamefont {Innocenti}},
  \bibinfo {author} {\bibfnamefont {A.}~\bibnamefont {Valletta}}, \ and\
  \bibinfo {author} {\bibfnamefont {A.}~\bibnamefont {Perali}},\ }\href
  {http://stacks.iop.org/1742-6596/529/i=1/a=012007} {\bibfield  {journal}
  {\bibinfo  {journal} {Journal of Physics: Conference Series}\ }\textbf
  {\bibinfo {volume} {529}},\ \bibinfo {pages} {012007} (\bibinfo {year}
  {2014})}\BibitemShut {NoStop}%
\bibitem [{\citenamefont {Shi}\ \emph {et~al.}(2016)\citenamefont {Shi},
  \citenamefont {Han}, \citenamefont {Peng}, \citenamefont {Richard},
  \citenamefont {Qian}, \citenamefont {Wu}, \citenamefont {Qiu}, \citenamefont
  {Wang}, \citenamefont {Hu}, \citenamefont {Y.-J.},\ and\ \citenamefont
  {Ding}}]{shi16}%
  \BibitemOpen
  \bibfield  {author} {\bibinfo {author} {\bibfnamefont {X.}~\bibnamefont
  {Shi}}, \bibinfo {author} {\bibfnamefont {Z.-Q.}\ \bibnamefont {Han}},
  \bibinfo {author} {\bibfnamefont {X.-L.}\ \bibnamefont {Peng}}, \bibinfo
  {author} {\bibfnamefont {R.}~\bibnamefont {Richard}}, \bibinfo {author}
  {\bibfnamefont {T.}~\bibnamefont {Qian}}, \bibinfo {author} {\bibfnamefont
  {X.-X.}\ \bibnamefont {Wu}}, \bibinfo {author} {\bibfnamefont {M.-W.}\
  \bibnamefont {Qiu}}, \bibinfo {author} {\bibfnamefont {S.~C.}\ \bibnamefont
  {Wang}}, \bibinfo {author} {\bibfnamefont {J.~P.}\ \bibnamefont {Hu}},
  \bibinfo {author} {\bibfnamefont {S.}~\bibnamefont {Y.-J.}}, \ and\ \bibinfo
  {author} {\bibfnamefont {H.}~\bibnamefont {Ding}},\ }\href
  {https://arxiv.org/abs/1606.01470} {\bibfield  {journal} {\bibinfo  {journal}
  {arXiv:1606.01470}\ } (\bibinfo {year} {2016})}\BibitemShut {NoStop}%
\bibitem [{\citenamefont {Perali}\ \emph {et~al.}(2012)\citenamefont {Perali},
  \citenamefont {Innocenti}, \citenamefont {Valletta},\ and\ \citenamefont
  {Bianconi}}]{perali12}%
  \BibitemOpen
  \bibfield  {author} {\bibinfo {author} {\bibfnamefont {A.}~\bibnamefont
  {Perali}}, \bibinfo {author} {\bibfnamefont {D.}~\bibnamefont {Innocenti}},
  \bibinfo {author} {\bibfnamefont {A.}~\bibnamefont {Valletta}}, \ and\
  \bibinfo {author} {\bibfnamefont {A.}~\bibnamefont {Bianconi}},\ }\href
  {http://stacks.iop.org/0953-2048/25/i=12/a=124002} {\bibfield  {journal}
  {\bibinfo  {journal} {Superconductor Science and Technology}\ }\textbf
  {\bibinfo {volume} {25}},\ \bibinfo {pages} {124002} (\bibinfo {year}
  {2012})}\BibitemShut {NoStop}%
\end{thebibliography}%
\end{document}